\documentclass[aps, prb, reprint, superscriptaddress]{revtex4-2}

\usepackage{graphicx} % no dvipdfmx
\usepackage{amsmath, amssymb, amsfonts, bm, mathrsfs, physics}
\usepackage{color}
\usepackage[T1]{fontenc}
\usepackage[normalem]{ulem}

\usepackage[colorlinks=true, linkcolor=blue, citecolor=red]{hyperref}

\begin{document}

\title{Hanbury Brown--Twiss interferometry at the $\nu=2/5$ fractional quantum Hall edge}

\author{Ryotaro Sano}
\email{r-sano@issp.u-tokyo.ac.jp}
\affiliation{Institute for Solid State Physics, The University of Tokyo, Kashiwa 277-8581, Japan}

\author{Fumihiro Murabayashi}
\affiliation{Institute for Solid State Physics, The University of Tokyo, Kashiwa 277-8581, Japan}

\author{Daigo Ichikawa}
\affiliation{Institute for Solid State Physics, The University of Tokyo, Kashiwa 277-8581, Japan}

\author{Thibaut Jonckheere}
\affiliation{Aix Marseille Univ, Universit\'e de Toulon, CNRS, CPT, Marseille, France}

\author{J\'er\^ome Rech}
\affiliation{Aix Marseille Univ, Universit\'e de Toulon, CNRS, CPT, Marseille, France}

\author{Thierry Martin}\affiliation{Aix Marseille Univ, Universit\'e de Toulon, CNRS, CPT, Marseille, France}
\author{Masayuki Hashisaka}
\affiliation{Institute for Solid State Physics, The University of Tokyo, Kashiwa 277-8581, Japan}

\author{Takeo Kato}
\affiliation{Institute for Solid State Physics, The University of Tokyo, Kashiwa 277-8581, Japan}

\begin{abstract}
We propose a Hanbury Brown--Twiss interferometer for a $\nu=2/5$ fractional quantum Hall edge system, in which quasiparticles tunnel between two co-propagating edge modes. In contrast {to} the previously studied anyonic Fabry--P\'{e}rot and Mach{--}Zehnder interferometers, the proposed setup relies purely on two-particle interference rather than single-particle interference. In the weak{-}tunneling regime, we employ a bosonized edge theory together with Keldysh perturbation theory to evaluate the cross-correlation of the tunneling currents. In the large-device limit, we obtain an analytic expression for the flux-dependent noise, whose structure closely resembles that of an electronic HBT interferometer{,} but with the electron charge replaced by the fractional charge $e^{\star}=e/3$ and with scaling dimensions characteristic of the fractional edge modes. In this limit, the explicit anyonic exchange phases cancel, whereas when the device size becomes comparable to the thermal length, the cross-correlation may recover a more explicit dependence on the anyonic statistical angle.
\end{abstract}

\maketitle

\section{Introduction}
    
The fractional quantum Hall effect (FQHE) provides one of the most celebrated arenas for emergent quasiparticles with fractional charge and anyonic exchange statistics~\cite{Laughlin1983PRL}. Since the original observation of plateaus at odd-denominator filling factors~\cite{Tsui1982PRL}, the existence of fractionally charged excitations has been established through tunneling shot{-}noise experiments~\cite{de-Picciotto1997Nature,Saminadayer1997PRL}. By contrast, the experimental detection of the underlying anyonic statistics {has} proved to be far more elusive; only in the past decade have interferometric measurements provided compelling evidence for Abelian braiding phases~\cite{Nakamura2020NatPhys}.

Most interferometer geometries studied to date fall into two categories: Fabry--P\'{e}rot (FP)~\cite{Chamon1997PRB,Halperin2011PRB,Nakamura2020NatPhys,Nakamura2022NatCommun,Nakamura2023PRX,Willett2023PRX,Kim2024NatNanotech,Ronetti2025PRL,Ronetti2025PRB,Samuelson2025arXiv} and Mach--Zehnder (MZ)~\cite{Ji2003Nature,Jonckheere2005PRB,Law2006PRB,Ponomarenko2007PRL,Feldman2007PRB,Kundu2023natphys,Batra2023PRB,Batra2025PRB,Ghosh2025NatPhys,Shimizu2025PRB} setups, in which single-particle amplitudes encircling an Aharonov--Bohm (AB) flux interfere. 
% ----
Recent experiments have also realized FP interference in bilayer graphene fractional quantum Hall states~\cite{Kim2024NatNanotech} and MZ interference in co-propagating integer edge channels~\cite{Shimizu2025PRB}.
% ---- Correspondence to the comment "If possible, I would appreciate it if you could also refer here to our recent paper on MZI in parallel integer edge states and Weizmann’s paper on graphene FPI. FPI: Kim, J., Dev, H., Kumar, R. et al. Aharonov–Bohm interference and statistical phase-jump evolution in fractional quantum Hall states in bilayer graphene. Nat. Nanotechnol. 19, 1619–1626 (2024). https://doi.org/10.1038/s41565-024-01751-w MZI: T. Shimizu et al., Mach–Zehnder interference of fractionalized electron-spin excitations, Phys. Rev. B 111, L161406 (2025). https://doi.org/10.1103/PhysRevB.111.L161406"
Although these single‑particle interferometers have yielded valuable insights into fractional charge and coherence, the observed phase shifts generally intertwine the effects of charge, exchange statistics, and edge structure, {thereby complicating} the unambiguous extraction of the statistical angle.

An alternative approach
is provided by two‑particle interferometry, pioneered in mesoscopic conductors through Hanbury Brown--Twiss (HBT) experiments~\cite{Henny1999Science,Oliver1999Science,Neder2007Nature}. In electronic HBT interferometer{s} the signal arises from exchange processes involving pairs of particles from independent sources, {without} single‑particle AB interference~\cite{Samuelsson2004PRL}. In the fractional quantum Hall context, {early work discussed the relation between partition noise and the statistics of tunneling excitations~\cite{Safi2001PRL}.} Campagnano \textit{et al.}~\cite{Campagnano2012PRL,Campagnano2013PRB} {have} proposed an anyonic HBT interferometer for a single‑mode edge, showing that the cross‑correlation of currents can reveal the fractional nature of the tunneling excitations even when single‑particle coherence is suppressed (see also Ref.~\cite{Vishveshwara2003PRL}). Closely related is the anyon collider~\cite{Rosenow2016PRL,Lee2022NatCommun,Bartolomei2020Science,Ruelle2023PRX,Glidic2023PRX,Schiller2023PRL,Lee2023Nature,Ruelle2025Science}, in which two independent quasiparticle beams collide and the resulting noise reveals exchange statistics without an enclosed AB loop. An intermediate {concept} %\tk{approach} 
between collider and interferometer geometries has also been explored recently in a FP setup, where stochastic anyon injection gives rise to time-domain braiding and AB oscillations in the current noise~\cite{girdhar2026arXiv}. Recent experiments at $\nu=1/3$~\cite{Bartolomei2020Science,Ruelle2023PRX,Glidic2023PRX} and $\nu=2/5$~\cite{Ruelle2023PRX,Glidic2023PRX} have confirmed that such two‑particle setups can probe anyonic statistics. However, the HBT concept has not yet been extended to hierarchical states with multiple co‑propagating edge modes. In hierarchical fractional quantum Hall states such as $\nu=2/5$, the edge structure naturally contains multiple co-propagating modes separated by incompressible strips. Tunneling between neighboring modes of the same chirality provides a natural platform for implementing HBT interferometry because such tunneling processes can be locally controlled by side gates without requiring counter-propagating channels or complex device geometries. A similar design principle was adopted in MZ setups for the $\nu=2/5$ Jain state~\cite{Batra2023PRB,Batra2025PRB}. Extending the HBT geometry to such hierarchical edges {thus} opens a new route to probing the properties of fractional quasiparticles in multi-mode edge systems.

\begin{figure}
\centering
\includegraphics[width=1.03\linewidth]{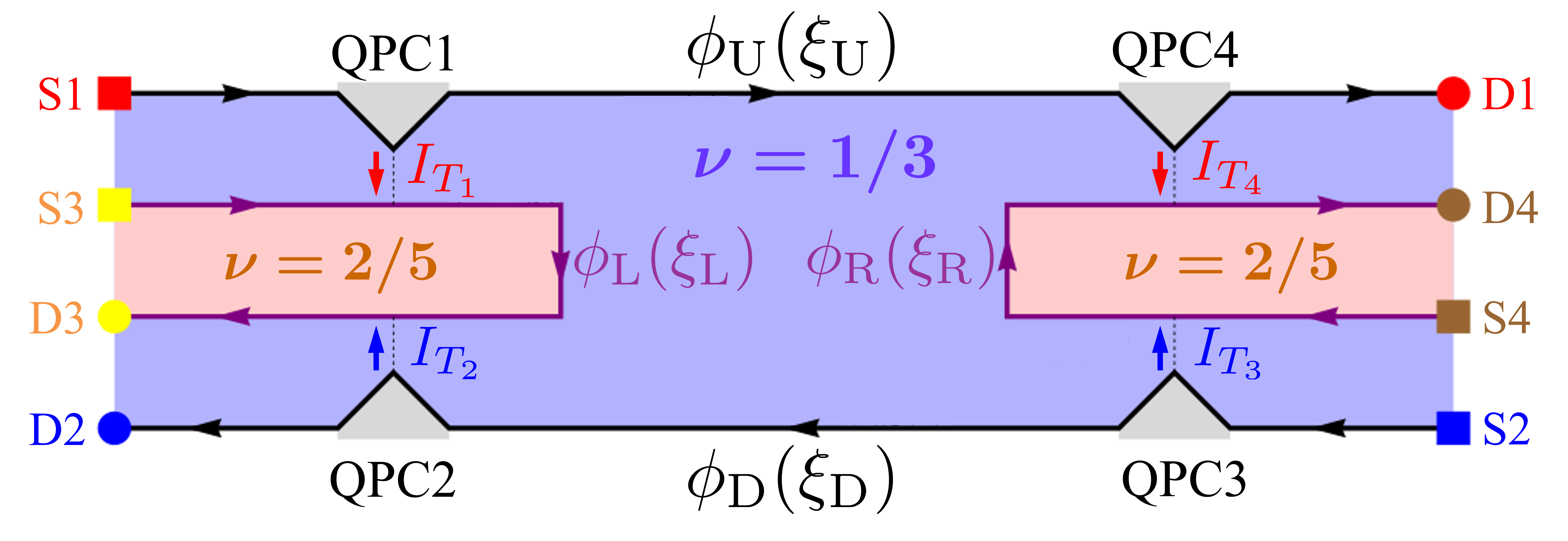}
\caption{Schematic of {a} Hanbury Brown{--}Twiss (HBT) interferometer in the $\nu=2/5$ fractional quantum Hall regime. Four QPCs locally bring adjacent edges {into close proximity facilitating} quasiparticle tunneling between the two co-propagating edge modes. The incoming edges are biased via Ohmic contacts S1 and S2 at dc voltages $V_{1}$ and $V_{2}$, whereas all other contacts (including S3 and S4) are {grounded}. The drain currents collected at D3 and D4 are used to form the cross-correlation noise $S_{34}$. Black and purple arrows indicate the chiral propagation direction{s}. Red and blue arrows specify the positive directions of the tunneling currents $I_{T_i}$ at each QPC.} 
\label{fig:device}
\end{figure}

Motivated by these considerations, we propose a Hanbury Brown--Twiss interferometer built on a $\nu=2/5$ fractional quantum Hall edge, as illustrated in Fig.~\ref{fig:device}. The edge consists of two co‑propagating modes separated by an incompressible $\nu=1/3$ strip, allowing quasiparticles with charge $e^{\star}=e/3$ to tunnel between the modes at a sequence of quantum point contacts (QPCs). In contrast to previous FP/MZ schemes, the flux-dependent signal in our geometry arises purely from two‑particle interference processes involving tunneling at all four QPCs. Within a bosonized description of the edge state{, using} Keldysh perturbation theory, we derive an {analytic} expression for the flux-dependent cross-correlation of the drain currents. In the large‑device limit, the resulting interference pattern takes a form analogous to an electronic HBT interferometer, with the electron charge replaced by $e^{\star}$ and with scaling dimensions characteristic of the fractional edge modes. A recent HBT proposal in a cross geometry for Laughlin states argued that the exchange phase can be extracted from the relative shift between the AB oscillations of a single-particle interference current and those of the current cross-correlation~\cite{puster2026arXiv}. In contrast, the present work addresses a multicomponent $\nu=2/5$ Jain edge, for which the flux-dependent cross-correlation exhibits a qualitatively different structure in the large-device regime.

The remainder of this paper is organized as follows. In Sec.~\ref{Sec:Formulation}, we set up the theoretical framework: bosonization of the chiral edge modes, the tunneling Hamiltonian with Klein factors, and the Keldysh perturbation theory used to evaluate the cross-correlation. {In Sec.~\ref{Sec:S34}, we derive} the flux-sensitive part of the cross-correlation in the large-device limit. {In Sec.~\ref{Sec:discussion}, we discuss} the physical implications of the result, including the roles of bias mismatch, temperature, arm mismatch, the flux-independent background, experimental considerations, and the regime beyond the large-device limit. {In Sec.~\ref{sec:summary}, we summarize} our main findings. Technical details concerning the enumeration of Keldysh contour orderings, the evaluation of the vertex-propagator Fourier transform, and the low-temperature reduction of the frequency kernel are deferred to Appendices~\ref{All-contour}, \ref{g_omega}, and \ref{Appendix:low-T_limit}, respectively.

\section{Model and method}
\label{Sec:Formulation}

We consider an edge of a topological liquid whose bulk realizes a fractional quantum Hall (FQH) state at filling factor $\nu=2/5$ (see Fig.~\ref{fig:device}). The edge structure consists of two co-propagating modes. We emphasize that the analysis does not rely on microscopic details of edge reconstruction but only on the presence of two co-propagating charge modes separated by an incompressible strip. Such a structure is expected for sufficiently smooth confinement potentials and has been widely used as an effective description of hierarchical $\nu=2/5$ edges. Using external gates, the two adjacent modes are {brought into close proximity} at several points, forming constrictions to facilitate quasiparticle tunneling. In this device geometry, the incompressible region separating the two modes has filling factor $\nu=1/3$, so that the relevant tunneling excitation across a constriction carries fractional charge $e^{\star}=e/3$. The interface between the $\nu=2/5$ liquid and the intervening $\nu=1/3$ strip supports an additional co-propagating edge mode with effective filling $\Delta\nu=2/5-1/3=1/15$. Multiple constrictions thereby define an interferometer geometry, which we model below. In close analogy with optical interferometry, each chiral edge channel plays the role of a waveguide{,} and each QPC acts as a beam splitter coupling two co‑propagating modes.

Within a bosonized description~\cite{Wen1990PRB,Wen1992IJMPB,Wen01101995,Kane2003PRB,Chang2003RMP}, we model the interferometer region by four chiral edge channels labeled by $\alpha \in\{\mathrm{U,D,L,R}\}$. Each channel is described by a chiral bosonic field $\phi_{\alpha}(\xi_{\alpha})$, where $\xi_{\alpha}$ is a coordinate along channel~$\alpha$ chosen to increase in the propagation direction. With this convention, all fields can be treated as right-moving, and the channel dependence enters only through the filling factors and mode velocities{:}
\begin{equation}
(\nu_{\alpha},\,v_{\alpha})=\left\{
\begin{array}{lc}
    (1/3,\,v_{1/3}),   & \alpha=\mathrm{U},\,\mathrm{D}, \\
   &                                 \\
    (1/15,\,v_{1/15}), & \alpha=\mathrm{L},\,\mathrm{R}.
\end{array}
\right.
\end{equation}
Accordingly, $\phi_{\mathrm{U}}$ and $\phi_{\mathrm{D}}$ ($\phi_{\mathrm{L}}$ and $\phi_{\mathrm{R}}$) describe chiral edge modes at $\nu=1/3$ ($\nu=1/15$) with velocity $v_{1/3}$ ($v_{1/15}$). The multichannel edge Hamiltonian reads
\begin{equation}
H_0 = \sum_{\alpha=\mathrm{U},\mathrm{D},\mathrm{L},\mathrm{R}}\frac{\hbar v_{\alpha}}{4\pi\nu_{\alpha}}        \int \dd{\xi_\alpha} \qty(\partial_{\xi_\alpha}\phi_{\alpha})^{2},\label{Eq:H_0}
\end{equation}
with {the} equal-time commutation relations
\begin{equation}
\qty[\phi_{\alpha}(\xi_{\alpha}),\phi_{\beta}(\xi'_{\beta})]=i\pi\nu_{\alpha} \delta_{\alpha\beta}\,\mathrm{sgn}(\xi_{\alpha}-\xi'_{\beta}),
\label{eq:bosoncommutation}
\end{equation}
so that fields on different channels commute. The charge density on channel $\alpha$ is given by
\begin{equation}
\rho_{\alpha}(\xi_{\alpha})=-\frac{e}{2\pi}\partial_{\xi_\alpha}\phi_{\alpha}(\xi_{\alpha}),
\end{equation}
and the total charge is given by $Q_{\alpha}=\int \dd{\xi_\alpha} \rho_{\alpha}$.

{To} investigate anyonic interference effects in the presence of multiple QPCs, we incorporate charge-conserving tunneling processes at each constriction. For notational convenience, we define $\phi_{\alpha}^{(i)} \equiv \phi_{\alpha}(\xi_{\alpha}^{(i)})$, where $\xi_{\alpha}^{(i)}$ is the position of the $i$th QPC along channel~$\alpha$. We consider four QPCs {that connect} the channel pairs $(\mathrm{U},\mathrm{L})$, $(\mathrm{L},\mathrm{D})$, $(\mathrm{D},\mathrm{R})$, and $(\mathrm{R},\mathrm{U})$ for $i=1,2,3,4$, respectively. For later convenience, we introduce the positive separation{s} between the {relevant} QPC positions along each channel, $L_{\mathrm{U}} \equiv \xi_{\mathrm{U}}^{(4)}-\xi_{\mathrm{U}}^{(1)}$, $L_{\mathrm{D}} \equiv \xi_{\mathrm{D}}^{(2)}-\xi_{\mathrm{D}}^{(3)}$, $L_{\mathrm{L}} \equiv \xi_{\mathrm{L}}^{(2)}-\xi_{\mathrm{L}}^{(1)}$, and $L_{\mathrm{R}} \equiv \xi_{\mathrm{R}}^{(4)}-\xi_{\mathrm{R}}^{(3)}$.
    
A local quasiparticle operator on channel $\alpha$ is represented by {a} vertex operator $e^{i\lambda_\alpha\phi_\alpha(\xi_\alpha)}$, where the integer $\lambda_{\alpha}$ specifies the quasiparticle type and carries electric charge $q_{\alpha}=-\lambda_{\alpha}\nu_{\alpha}e$ as follows from {the commutation relation} $[Q_{\alpha},e^{i\lambda_\alpha\phi_\alpha(\xi_\alpha)}]=q_{\alpha}e^{i\lambda_\alpha\phi_\alpha(\xi_\alpha)}$. Charge conservation at each QPC requires the tunneling operator to be electrically neutral. {This charge neutrality, together with the channel fillings $\nu_{\mathrm{U,D}}=1/3$ and $\nu_{\mathrm{L,R}}=1/15$, implies $\lambda_{\mathrm{L,R}} = 5\lambda_{\mathrm{U,D}}$.}
The leading process corresponds to $\lambda_{\mathrm{U,D}}=1$ and hence $\lambda_{\mathrm{L,R}}=5$, transfers {a} charge $e^{\star}=e/3$ across the constriction, and yields the bosonic {operator} structure $e^{-5i\phi_\mathrm{L,R}}e^{+i\phi_\mathrm{U,D}}$ together with its Hermitian conjugate{,} describing the reverse process. Accordingly, the tunneling Hamiltonian $H_{T}$ is given by
\begin{equation}
H_{T}=T_{1}+T_{2}+T_{3}+T_{4},
\end{equation}
with $T_{i}=A_{i}+A_{i}^{\dagger}$, where $A_{i}$ takes the form
\begin{subequations}
\label{Eq:A_i}
\begin{align}
    A_{1} & =\frac{\Gamma_{1}\mathcal{F}_{1}}{2\pi\sqrt{a_{1/3}a_{1/15}}}e^{-5i\phi_\mathrm{L}^{(1)}}e^{+i\phi_\mathrm{U}^{(1)}}, \\
    A_{2} & =\frac{\Gamma_{2}\mathcal{F}_{2}}{2\pi\sqrt{a_{1/3}a_{1/15}}}e^{-5i\phi_\mathrm{L}^{(2)}}e^{+i\phi_\mathrm{D}^{(2)}}, \\
    A_{3} & =\frac{\Gamma_{3}\mathcal{F}_{3}}{2\pi\sqrt{a_{1/3}a_{1/15}}}e^{-5i\phi_\mathrm{R}^{(3)}}e^{+i\phi_\mathrm{D}^{(3)}}, \\
    A_{4} & =\frac{\Gamma_{4}\mathcal{F}_{4}}{2\pi\sqrt{a_{1/3}a_{1/15}}}e^{-5i\phi_\mathrm{R}^{(4)}}e^{+i\phi_\mathrm{U}^{(4)}}.
\end{align}
\end{subequations}
Here, $a_{1/3}$ and $a_{1/15}$ are short-distance ({ultraviolet}) cutoffs and $\Gamma_{i}$ are complex tunneling amplitudes. 
    
For {systems with} multiple QPCs, the bosonic vertex operators alone do not fully determine the mutual statistics of tunneling events at distinct {QPCs}. {Since} tunneling operators at different QPCs may involve bosonic fields associated with 
a shared edge segment, the bosonic vertex parts can acquire nontrivial exchange phases and therefore {do} not commute. We thus attach Klein factors $\mathcal{F}_{i}$ to each $A_{i}$ and choose their algebra so that the full tunneling {operators} commute at equal times, $[T_{i}(t),T_{j}(t)]=0$ for $i\neq j$. This equal-time commutativity is the locality requirement for spatially separated tunneling terms in the effective low-energy Hamiltonian~\cite{Campagnano2013PRB}. A convenient implementation is to endow the Klein factors with a cyclic exchange algebra. Since the QPCs are labeled sequentially in counterclockwise order around the perimeter of the interferometer, as shown in Fig.~\ref{fig:device}, only neighboring tunneling operators share an edge segment, {whereas} non-neighboring operators act on disjoint field sets and hence commute in their bosonic parts. Consequently, it {suffices to} impose nontrivial exchange relations only between neighboring Klein factors, together with the cyclic identification $\mathcal{F}_{i+4}\equiv\mathcal{F}_{i}$: explicitly, we take
\begin{subequations}
\label{Eq:Klein-algebraset}
\begin{align}
\label{Eq:Klein-algebra}
\mathcal{F}_{i}\mathcal{F}_{i+1}e^{-i\pi/3} & =\mathcal{F}_{i+1}\mathcal{F}_{i}, \\            \mathcal{F}_{i}\mathcal{F}_{i\pm1}^{\dagger}e^{\pm i\pi/3}
& =\mathcal{F}_{i\pm1}^{\dagger}\mathcal{F}_{i},
\label{Eq:Klein-algebra2}
\end{align}
\end{subequations}
while all non-neighboring pairs commute. The exchange phases of the bosonic vertex parts follow from the Baker-Campbell{--}Hausdorff identity $e^{A}e^{B}=e^{B}e^{A}e^{[A,B]}$, since the exponents are linear in chiral bosonic fields and their commutators are $c$-numbers.
The tunneling current through the {$i$th} QPC is defined as the net rate at which a quasiparticle of charge $e^{\star}$ is transferred across the constriction. The {corresponding} current operator is then
\begin{equation}
I_{T_i}=i\frac{e^{\star}}{\hbar}\bigl(A_{i}-A_{i}^{\dagger}\bigr),
\end{equation}
where the sign {convention} $I_{T_i}>0$ is fixed to correspond to quasiparticle transfer along the arrow direction assigned to {QPC$i$ ($i=1,2,3,4$)} in Fig.~\ref{fig:device}.

We now turn to the stationary nonequilibrium transport and {current} noise. Assuming that $H_{T}$ is switched on adiabatically from $t=-\infty$, we evaluate the drain currents and their cross-correlation within the Keldysh {formalism}~\cite{Keldysh1964} {(see also Ref.~\cite{martin05} for a review of mesoscopic noise)}. {The dc voltages $V_{1}$ and $V_{2}$ are} applied to the incoming U and D {channels} via the source contacts S1 and S2, respectively, whereas the downstream drains D3 and D4 are {grounded}. With these biases, we take $I_{T_i}>0$ to correspond to quasiparticle transfer from the biased incoming edges U and D, consistent with the arrows in Fig.~\ref{fig:device}. The current measured at each drain is therefore given by the total tunneling current feeding that drain. With the sign {convention} indicated in Fig.~\ref{fig:device}, we define
\begin{subequations}
\label{Eq:D_currents}
\begin{align}
I_{\mathrm{D3}}\equiv I_{T_1}+I_{T_2}=i\frac{e^{\star}}{\hbar}\bigl(A_{1}-A_{1}^{\dagger}+A_{2}-A_{2}^{\dagger}\bigr), \\
I_{\mathrm{D4}}\equiv I_{T_3}+I_{T_4}=i\frac{e^{\star}}{\hbar}\bigl(A_{3}-A_{3}^{\dagger}+A_{4}-A_{4}^{\dagger}\bigr).
\end{align}
\end{subequations}
In the bosonized description, the voltage biases can be incorporated by attaching dynamical phases to quasiparticle operators on the biased incoming edges. Explicitly, we perform the replacements
\begin{subequations}
\begin{gather}
e^{i\phi_\mathrm{U}(\xi_\mathrm{U},t)}\to e^{i\phi_\mathrm{U}(\xi_\mathrm{U},t)}
e^{i(e^\star V_1/\hbar)(t-\xi_\mathrm{U}/v_{1/3})},\\
e^{i\phi_\mathrm{D}(\xi_\mathrm{D},t)}\to e^{i\phi_\mathrm{D}(\xi_\mathrm{D},t)}
e^{i(e^\star V_2/\hbar)(t-\xi_\mathrm{D}/v_{1/3})}.
\end{gather}
\end{subequations}
Throughout, {the} operator time dependence is taken in the interaction picture with respect to $H_{0}$, while the dc biases enter only through the above $c$-number phases attached to $A_{i}(t)$. Consequently, $A_{1}$ and $A_{4}$ acquire the phase associated with $V_{1}$, whereas $A_{2}$ and $A_{3}$ acquire that associated with $V_{2}$~\cite{Law2006PRB}. The phase factor can be absorbed in the tunneling amplitude{s}
\begin{subequations}
\label{Eq:bias-shift}
\begin{align}
\Gamma_{1}\to\Gamma_{1}\,e^{+i(e^\star V_1/\hbar)(t-\xi_\mathrm{U}^{(1)}/v_{1/3})}, \\
\Gamma_{2}\to\Gamma_{2}\,e^{+i(e^\star V_2/\hbar)(t-\xi_\mathrm{D}^{(2)}/v_{1/3})}, \\
\Gamma_{3}\to\Gamma_{3}\,e^{+i(e^\star V_2/\hbar)(t-\xi_\mathrm{D}^{(3)}/v_{1/3})}, \\
\Gamma_{4}\to\Gamma_{4}\,e^{+i(e^\star V_1/\hbar)(t-\xi_\mathrm{U}^{(4)}/v_{1/3})},
\end{align}
\end{subequations}

We focus on the zero-frequency cross-correlation between the drain currents, {defined as}
\begin{equation}
S_{34}=\frac{1}{2}\sum_{\eta}\int_{-\infty}^{+\infty}\dd{(t-t')}\expval{\mathcal{T_C}I_\mathrm{D3}(t_\eta)I_\mathrm{D4}(t'_{-\eta})\mathcal{S_C}}_{0},\label{def:S_34}
\end{equation}
with
\begin{equation}
\mathcal{S_C}=\mathcal{T_C}\exp\biggl[-\frac{i}{\hbar}\sum_{\eta'}\int_{-\infty}^{+\infty}\dd{t'_{\eta'}}H_{T}(t'_{\eta'})\biggr].
\end{equation}
Here, $\mathcal{T_C}$ {denotes the contour time-ordering} and $\eta,\eta'=\pm$ label the forward {and} backward branches of the Keldysh contour, respectively~\cite{Keldysh1964}.

In the interaction picture, the expectation values in Eq.~\eqref{def:S_34} are expressed in terms of contour-ordered correlators of quasiparticle vertex operators. A key ingredient is the contour-ordered Gaussian vertex correlator{, given by}
\begin{align}
& \expval{\mathcal{T_C}\,e^{i{\lambda}\phi_{\alpha}(\xi,t_\eta)}e^{i{\lambda'}\phi_\beta(\xi',t'_{\eta'})}}_{0}=\delta_{\alpha\beta}\delta_{\lambda+\lambda',0}\,\mathcal{G}^{\eta\eta'}_{\alpha}(\tau_{\alpha}),\label{Eq:vertex}
\end{align}
where $\tau_{\alpha}\equiv t-t'-(\xi-\xi')/v_{\alpha}$, and charge neutrality within a given channel enforces $\lambda'=-\lambda$. Here, $\mathcal{G}^{\eta\eta'}_{\alpha}$ is the {ultraviolet}-regularized vertex propagator on channel $\alpha$. We adopt the explicit finite-temperature form
\begin{equation}
\mathcal{G}^{\eta\eta'}_{\alpha}(\tau_{\alpha})=\left[\frac{\sinh\frac{i\pi\tau_{0}}{\beta\hbar}}{\sinh\frac{\pi}{\beta\hbar}\left(i\tau_{0}-\sigma_{tt'}^{\eta\eta'}\tau_{\alpha}\right)}
\right]^{\lambda_\alpha^2\nu_\alpha},
\end{equation}
where the {ultraviolet} cutoff is implemented by a short-time regulator $\tau_{0}$ with $a_{\alpha}\equiv v_{\alpha}\tau_{0}$. Here,
\begin{equation}
\sigma_{tt'}^{\eta\eta'}\equiv\frac{\eta+\eta'}{2}\mathrm{sgn}(t-t')+\frac{\eta'-\eta}{2},
\label{Eq:sigma}
\end{equation}
encodes contour {time-}ordering on the Keldysh branches $\eta,\eta'=\pm$. {Since} the unperturbed Hamiltonian $H_{0}$ in Eq.~\eqref{Eq:H_0} is a direct sum of independent channels and bosonic fields on different channels commute, mixed-channel contractions vanish and multipoint correlators factorize into products of single-channel correlators. With the bias phases absorbed into $A_{i}(t)$, all voltage dependence enters through these $c$-number phases, while the remaining dynamical input is encoded in $\mathcal{G}_{\alpha}^{\eta\eta'}$. Expanding $\mathcal{S_C}$ to the lowest nonvanishing order in the tunneling amplitudes $\Gamma_{i}$, the resulting expressions involve only products of such vertex correlators.

\section{Flux-sensitive cross-correlation}
\label{Sec:S34}

Since each drain current in Eq.~\eqref{Eq:D_currents} is linear in $A_{i}$ and $A_{i}^{\dagger}$, the lowest nonvanishing contribution to $S_{34}$ is obtained by expanding $\mathcal{S_C}$ to second order in $H_{T}$. To this lowest nontrivial order, we {obtain}
\begin{widetext}
\begin{equation}
S_{34} = \frac{(-i)^{2}}{2\cdot2!\hbar^{2}}
\sum_{\eta,\eta_1,\eta_2}\eta_{1}\eta_{2}
\int_{-\infty}^{+\infty}\dd{(t-t')}\int_{-\infty}^{+\infty}\dd{t_1}
\int_{-\infty}^{+\infty}\dd{t_2}
\expval{\mathcal{T_C}I_\mathrm{D3}(t_\eta)I_\mathrm{D4}(t'_{-\eta})H_T(t_{1,\eta_1})H_T(t_{2,\eta_2})}_{0}.
\label{Eq:S34}
\end{equation}
Among the contributions generated from Eq.~\eqref{Eq:S34}, the Aharonov{--}Bohm (AB) flux enter{s} only through interference processes that effectively enclose the interferometer area. In the present four-QPC geometry, the minimal closed loop at this order necessarily involves tunneling at all four QPCs and is proportional to $\Gamma_{1}\Gamma_{2}^{\ast}\Gamma_{3}\Gamma_{4}^{\ast}$ together with its complex conjugate. All other terms at this order involve at most two QPCs and are flux independent. We therefore isolate the AB-flux-dependent part of the cross-correlation by retaining only those contractions involving all four QPCs. We denote this flux-dependent contribution by $S_{34}^{\Phi}$. A representative operator structure is $A_{2}^{\dagger}(t_{\eta})A_{4}^{\dagger}(t'_{-\eta})A_{1}(t_{1,\eta_1}) A_{3}(t_{2,\eta_2})$, which yields
\begin{equation}
\left(\frac{e^{\star}}{2\hbar^{2}}\right)^{2}\sum_{\eta,\eta_1,\eta_2}
\eta_{1}\eta_{2}\int_{-\infty}^{+\infty}\dd{(t-t')}\int_{-\infty}^{+\infty}\dd{t_1}
\int_{-\infty}^{+\infty} \dd{t_2} \expval{\mathcal{T_C}A_2^\dagger(t_\eta)A_4^\dagger(t'_{-\eta})A_1(t_{1,\eta_1})A_3(t_{2,\eta_2})}_{0}.
\end{equation}
Substituting Eq.~\eqref{Eq:A_i} and using the shorthand $\phi_{\alpha}^{(i)}(t_{\eta})\equiv\phi_{\alpha}(\xi_{\alpha}^{(i)},t_{\eta})$, the contour-ordered expectation value factorizes into a Klein-factor part and a bosonic vertex part. The representative contribution evaluates to
\begin{align}
& \frac{\Gamma_{1}\Gamma_{2}^{\ast}\Gamma_{3}\Gamma_{4}^{\ast}\,e^{2\pi i\Phi/\Phi_0^\star}}{(2\pi)^{4}(a_{1/3}a_{1/15})^{2}}\mathcal{F}_{2}^{\dagger}\mathcal{F}_{4}^{\dagger}\mathcal{F}_{1}\mathcal{F}_{3}\,e^{i(e^\star V_1/\hbar)(t_1-t'+L_\mathrm{U}/v_{1/3})}\,e^{i(e^\star V_2/\hbar)(t_2-t+L_\mathrm{D}/v_{1/3})}\nonumber                   \\
& \times\expval{\mathcal{T_C}e^{-i\phi_\mathrm{D}^{(2)}(t_\eta)}e^{+5i\phi_\mathrm{L}^{(2)}(t_\eta)}e^{-i\phi_\mathrm{U}^{(4)}(t'_{-\eta})}e^{+5i\phi_\mathrm{R}^{(4)}(t'_{-\eta})}e^{-5i\phi_\mathrm{L}^{(1)}(t_{1,\eta_1})}e^{+i\phi_\mathrm{U}^{(1)}(t_{1,\eta_1})}e^{-5i\phi_\mathrm{R}^{(3)}(t_{2,\eta_2})}e^{+i\phi_\mathrm{D}^{(3)}(t_{2,\eta_2})}}_{0}.
\end{align}
The AB phase $2\pi\Phi/\Phi_0^{\star}$ enters through the gauge-invariant phase of the tunneling amplitude with the fractional flux quantum $\Phi_0^{\star}=h/e^{\star}$. Since $H_{0}$ is quadratic, the bosonic vertex part {can be} evaluated exactly using Eq.~\eqref{Eq:vertex} as
\begin{align}
& \frac{\Gamma_{1}\Gamma_{2}^{\ast}\Gamma_{3}\Gamma_{4}^{\ast}\,e^{2\pi i\Phi/\Phi_0^\star}}{(2\pi)^{4}(a_{1/3}a_{1/15})^{2}}\mathcal{F}_{2}^{\dagger}\mathcal{F}_{4}^{\dagger}\mathcal{F}_{1}\mathcal{F}_{3}\,e^{i(e^\star V_1/\hbar)(t_1-t'+L_\mathrm{U}/v_{1/3})}\,e^{i(e^\star V_2/\hbar)(t_2-t+L_\mathrm{D}/v_{1/3})}\nonumber \\
& \times\mathcal{G}_{\mathrm{D}}^{\eta\eta_2}(t-t_{2}-L_{\mathrm{D}}/v_{1/3})\mathcal{G}_{\mathrm{L}}^{\eta\eta_1}(t-t_{1}-L_{\mathrm{L}}/v_{1/15})\mathcal{G}_{\mathrm{U}}^{-\eta\eta_1}(t'-t_{1}-L_{\mathrm{U}}/v_{1/3})\mathcal{G}_{\mathrm{R}}^{-\eta\eta_2}(t'-t_{2}-L_{\mathrm{R}}/v_{1/15}).\label{2413}
\end{align}
\end{widetext}
The remaining AB{-}phase-sensitive contributions are obtained from the other contour orderings of the same four tunneling operators inside $\mathcal{T_C}$. Under our cyclic Klein-factor algebra, the interchange $(t_{1},\eta_{1})\leftrightarrow (t_{2},\eta_{2})$ is phase-free because it only exchanges the non-neighboring Klein factors $\mathcal{F}_{1}$ and $\mathcal{F}_{3}$, whereas the other contour orderings may involve exchanges of neighboring Klein factors and therefore generate anyonic statistical phases. We do not list all AB-sensitive contour orderings explicitly in the main text; the full set of inequivalent orderings and their associated Klein-factor products are summarized in Appendix~\ref{All-contour}.

% \color {blue} {Question to Sano-san: do you think it is possible to include this kind of physical, non-technical argument (in addition to more precise explanations already given) ?\textit{Physically, if the distance between the first and second QPC on each bosonic edge is large, then the tunneling at the first QPCs (top left and bottom right on figure 1) occurs much before the tunneling at the second QPCs (bottom left and top right), and thus the tunneling operators of the first QPCs are always at earlier times than those at the last QPCs, and we don’t need to worry about the Klein factor phase due tothe commutation of these operators.}} \red{I fully agree with your suggestion and included in the Discussion. See [\#1]} \color{black}

A crucial simplification occurs in the large-device limit, $L_\alpha/v_\alpha\gg\beta\hbar$~\cite{Campagnano2013PRB}. In this regime, each finite-temperature propagator $\mathcal{G}_{\alpha}^{\eta\eta'}(\tau_{\alpha})$ in Eq.~\eqref{2413} is sharply localized within a temporal window of width $\sim\beta\hbar$ around $\tau_{\alpha}=0$ and is exponentially suppressed outside this window. For the surviving four-QPC contribution, the four propagators therefore constrain the time {integrals} to a common region in which the operator insertions are separated by the positive time of flight between neighboring QPCs. Since each $\tau_{\alpha}$ is defined by the retarded-time combination of the form $t-t'-L_{\alpha}/v_{\alpha}$, the dominant contribution comes from the region in which each propagator is evaluated near its retarded-time peak, satisfying $\abs{\tau_\alpha}\lesssim\beta\hbar$. With our convention $L_{\alpha}>0$, this implies $t-t'>0$ throughout the dominant region, so that $\mathrm{sgn}(t-t')=+1$. Substituting this into Eq.~\eqref{Eq:sigma}, we obtain $\sigma_{tt'}^{\eta\eta'}=\eta'$. The vertex propagator therefore becomes independent of {the} first Keldysh index. As a result, when the weighted Keldysh sums $\sum_{\eta_1=\pm}\eta_{1}$ and $\sum_{\eta_2=\pm}\eta_{2}$ are carried out, contributions in which $\eta_{1}$ or $\eta_{2}$ enter only through the first indices of the propagators cancel pairwise. In our four-QPC problem, this cancellation eliminates all contour orderings except the pair $A_{2}^{\dagger}(t_{\eta})A_{4}^{\dagger}(t'_{-\eta})A_{1}(t_{1,\eta_1})A_{3}(t_{2,\eta_2})$ and $A_{2}^{\dagger}(t_{\eta})A_{4}^{\dagger}(t'_{-\eta})A_{3}(t_{1,\eta_1})A_{1}(t_{2,\eta_2})$.
We therefore restrict to this surviving pair hereafter. For these terms it is convenient to change integration variables to $\tilde{t}_{1}=t_{1}-t'+L_{\mathrm{U}}/v_{1/3}$ and $\tilde{t}_{2}=t_{2}-t+L_{\mathrm{D}}/v_{1/3}$, which casts the remaining time integrals into the following convolution:
\begin{widetext}
\begin{align}
& 2\sum_{\eta,\eta_1,\eta_2}\eta_{1}\eta_{2}\int_{-\infty}^{+\infty}\dd{(t-t')}\int_{-\infty}^{+\infty}\dd{\tilde{t}_1}\int_{-\infty}^{+\infty}\dd{\tilde{t}_2}\,e^{i(e^\star V_1/\hbar)\tilde{t}_1}\,e^{i(e^\star V_2/\hbar)\tilde{t}_2}\nonumber  \\
& \qquad\times\mathcal{G}_{\mathrm{D}}^{\eta\eta_2}(-\tilde{t}_{2})\mathcal{G}_{\mathrm{L}}^{\eta\eta_1}(t-t'-\tilde{t}_{1}+L_{\mathrm{U}}/v_{1/3}-L_{\mathrm{L}}/v_{1/15})\mathcal{G}_{\mathrm{U}}^{-\eta\eta_1}(-\tilde{t}_{1})\mathcal{G}_{\mathrm{R}}^{-\eta\eta_2}(t'-t-\tilde{t}_{2}+L_{\mathrm{D}}/v_{1/3}-L_{\mathrm{R}}/v_{1/15})\nonumber \\
& =2\sum_{\eta,\eta_1,\eta_2}\eta_{1}\eta_{2}\int_{-\infty}^{+\infty}\frac{\dd{\omega}}{2\pi}g_{\mathrm{D}}^{\eta\eta_2}\left(\omega-\frac{e^{\star}V_{2}}{\hbar}\right)g_{\mathrm{L}}^{\eta\eta_1}(-\omega)g_{\mathrm{U}}^{-\eta\eta_1}\left(\omega-\frac{e^{\star}V_{1}}{\hbar}\right)g_{\mathrm{R}}^{-\eta\eta_2}(-\omega)\,e^{i\omega\Delta\tau},\label{Eq:gggg}
\end{align}
\end{widetext}
where we have introduced the Fourier transform of the vertex propagator
\begin{equation}
{g}_{\mathrm{\alpha}}^{\eta\eta'}(\omega)\equiv\int_{-\infty}^{+\infty}\dd
{\tau}e^{i\omega\tau}\mathcal{G}_{\alpha}^{\eta\eta'}(\tau),\label{def:Fourier}
\end{equation}
and the net propagation-time mismatch
\begin{equation}
\Delta\tau\equiv\frac{L_{\mathrm{U}}+L_{\mathrm{D}}}{v_{1/3}}-\frac{L_{\mathrm{L}}+L_{\mathrm{R}}}{v_{1/15}} ,
\end{equation}
which is the relative delay accumulated along the two arms. 
The two surviving contour orderings give identical contributions, yielding the overall factor of 2 in Eq.~\eqref{Eq:gggg}. The explicit form of $g_{\alpha}^{\eta\eta'}(\omega )$ is given by (see Appendix~\ref{g_omega} for the derivation)
\begin{align}
g_{\alpha}^{\eta\eta'}(\omega) & =\frac{\beta\hbar}{2\pi}\left(\frac{2\pi\tau_{0}}{\beta\hbar}\right)^{\lambda_\alpha^2\nu_\alpha}e^{\eta'\beta\hbar\omega/2}\nonumber                                                      \\
& \quad\times B\left(\frac{\lambda_{\alpha}^{2}\nu_{\alpha}}{2}-i\frac{\beta\hbar\omega}{2\pi},\frac{\lambda_{\alpha}^{2}\nu_{\alpha}}{2}+i\frac{\beta\hbar\omega}{2\pi}\right),\label{Eq:g}
\end{align}
where $B(x,y)\equiv\Gamma(x)\Gamma(y)/\Gamma(x+y)$ is the Euler beta function with the Euler gamma function $\Gamma(z)$. Finally, substituting this into Eq.~\eqref{Eq:gggg} and summing up all the Keldysh indices, we arrive at a compact analytical form:
\begin{widetext}
\begin{align}
S_{34}^{\Phi} & =\frac{4e^{\star2}}{(2\pi)^{4}}\frac{\Gamma_{1}\Gamma_{2}^{\ast}\Gamma_{3}\Gamma_{4}^{\ast}\,e^{2\pi i\Phi/\Phi_0^\star}}{\hbar^{4}v_{1/3}^{2}v_{1/15}^{2}\Gamma^{2}(5/3)\Gamma^{2}(1/3)}\sinh(\frac{\beta e^{\star}V_{1}}{2})\sinh(\frac{\beta e^{\star}V_{2}}{2})\nonumber\\
& \quad\times\int_{-\infty}^{+\infty}\frac{\dd{\omega}}{2\pi}e^{i\omega\Delta\tau}\abs{\Gamma\left(\frac{5}{6}-i\frac{\beta\hbar\omega}{2\pi}\right)}^{4}\abs{\Gamma\left(\frac{1}{6}-i\frac{\beta(e^{\star}V_{1}-\hbar\omega)}{2\pi}\right)}^{2}\abs{\Gamma\left(\frac{1}{6}-i\frac{\beta(e^{\star}V_{2}-\hbar\omega)}{2\pi}\right)}^{2}+\mathrm{c.c.}.\label{Result:S_34}
\end{align}
\end{widetext}
Eq.~\eqref{Result:S_34} is the central result of this work. In the large{-}device limit considered here, the surviving flux-dependent contribution has the same overall structure as in an electronic HBT interferometer~\cite{Campagnano2013PRB}, but with the electron charge replaced by the fractional quasiparticle charge $e^{\star}=e/3$ and with scaling dimensions of the tunneling operators. Notably, the explicit statistical phase factors that arise in intermediate contour orderings cancel out in this limit; the remaining signal therefore reflects the fractional charge and scaling dimensions of the tunneling quasiparticles, rather than {the} anyonic statistical angle.

\section{Discussion}
\label{Sec:discussion}

\begin{figure}[t]
\centering
\includegraphics[width=0.99\linewidth]{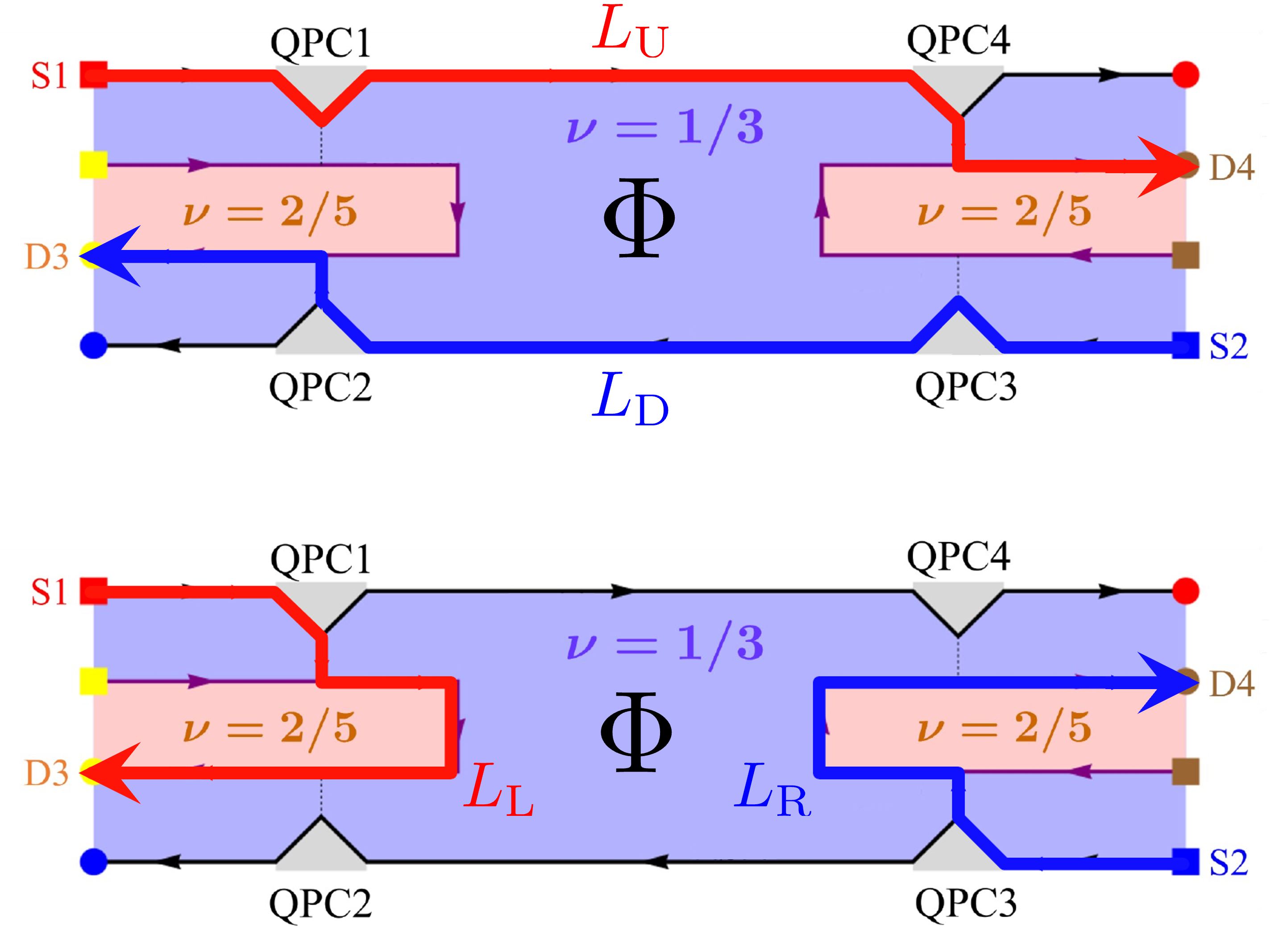}
\caption{
Schematic of two interfering two-particle paths contributing to $S_{34}^\Phi$. In each configuration, one excitation is injected from source S1 and the other from source S2, and the two excitations exit into drains D3 and D4. The relative phase between the two paths contains the AB phase associated with the enclosed {magnetic} flux $\Phi$. The segment lengths $L_\mathrm{U}$, $L_\mathrm{D}$, $L_\mathrm{L}$, and $L_\mathrm{R}$ determine the propagation times and thereby the arm-mismatch delay $\Delta\tau$ between the two interfering paths.}
\label{fig:interference}
\end{figure}

Eq.~\eqref{Result:S_34} admits a simple physical interpretation, {as} summarized schematically in Fig.~\ref{fig:interference}. The AB-sensitive part of the cross-correlation arises from the interference between two distinct two-particle path{s}, in which one excitation injected from S1 and the other from S2 are collected at D3 and D4 through different sequences of tunneling and propagation. 
In this sense, the present setup realizes a fractional analogue of an HBT interferometer in a hierarchical $\nu=2/5$ edge structure. 

% Added discussion paragraph corresponding to the comment #1
The large-device limit also admits a simple physical interpretation. When the times of flight between the relevant QPCs along each chiral edge are much longer than the thermal time scale, the interference process is effectively resolved into two temporally separated stages: tunneling at the upstream QPCs occurs well before tunneling at the downstream QPCs. The relevant histories are then no longer related by a pure exchange of indistinguishable quasiparticles, but instead correspond to sequential, causally ordered tunneling events. As a result, the anyonic exchange phase does not appear as a simple additive {phase} shift of the AB oscillation{s} in this regime. This {can} be contrasted with the recent single-mode Laughlin HBT proposal~\cite{puster2026arXiv}, where the interfering two-particle processes are {more directly related} by quasiparticle exchange, leading to an additive exchange phase shift relative to the current oscillations. In the present multicomponent $\nu=2/5$ Jain-edge setup, by contrast, the AB-sensitive contribution does not reduce to such a shifted form in the large-device limit.
    
Within the large-device limit {considered} here, the surviving interference is controlled primarily by three ingredients: the quasiparticle charge $e^{\star}$, the scaling dimensions of the tunneling operators, and the propagation-time mismatch $\Delta\tau$ between the two interfering paths. At the same time, the weighted Keldysh sums show that the contour orderings carrying explicit anyonic exchange phases cancel in this regime. Accordingly, the flux dependence does not isolate the statistical angle as an independent phase shift. Instead, the fractional nature of the signal is encoded in the charge $e^{\star}= e/3$ and in the nontrivial scaling dimensions entering Eq.~\eqref{Result:S_34}. {This behavior} reflects the temporal separation of tunneling events imposed by the large-device limit, rather than a limitation of the HBT geometry itself. Outside this regime, where the propagation time between QPCs becomes comparable to the thermal coherence time, the cancellation mechanism may no longer operate and the cross-correlation may retain a more explicit dependence on the anyonic statistical angle.

\subsection{Low-temperature asymptotics within the large-device limit}

\begin{figure}[t]
\centering
\includegraphics[width=0.99\linewidth]{S34_delta.png}
\caption{Flux-dependent cross-correlation as a function of the bias mismatch ratio $r=V_</V_>$ at $\Delta\tau=0$. The main panel shows $S_{34}^{\Phi}/V_{>}$ for several values of $k_\mathrm{B}T/e^\star V_>$, together with the zero-temperature asymptotic result (dashed line). The inset shows the small-$r$ region on a log-log scale; the {guideline} indicates the asymptotic scaling $S_{34}^{\Phi}/V_{>}\propto r^{5/3}$. An overall prefactor $2(e^{\star3}/\hbar^5v_{1/3}^2v_{1/15}^2)\Re\,[\Gamma_1\Gamma_2^\ast\Gamma_3\Gamma_4^\ast e^{2\pi i\Phi/\Phi_0^\star}]$ {that is} common to all curves has been omitted.}
\label{fig:S34_ratio}
\end{figure}

Strictly speaking, the fixed-geometry zero-temperature limit is not compatible with the large-device condition $L_\alpha/v_\alpha\gg\beta\hbar$, since $\beta\to\infty$ as $T\to0$. The analysis below should therefore be interpreted as the leading low-temperature asymptotics within the large-device regime, or equivalently as a simultaneous limit in which the edge-mode times of flight $L_\alpha/v_\alpha$ remain the longest time scales. In this regime, the asymptotic form of the frequency kernel in Eq.~\eqref{Result:S_34} shows that the leading contribution survives only in the overlap of the two bias windows. Consequently, the interference signal is exponentially suppressed for opposite-sign biases, whereas for $V_1V_2>0$ it is controlled by the common frequency interval selected by the two voltages. This makes the low-temperature signal highly sensitive to the bias mismatch and explains its enhancement near the symmetric-bias configuration.
For $\Delta\tau=0$, the surviving finite-interval power-law integral can be evaluated in closed form, and both sign choices reduce to the same Gauss hypergeometric expression,
\begin{align}
S_{34}^\Phi&=\frac{e^{\star3}}{2\pi\hbar}\frac{[\Gamma_{1}\Gamma_{2}^{\ast}\Gamma_{3}\Gamma_{4}^{\ast}\,e^{2\pi i\Phi/\Phi_0^\star}}{\hbar^{4}v_{1/3}^{2}v_{1/15}^{2}\Gamma^2(2/3)\Gamma(8/3)}\nonumber\\
&\qquad\times\frac{V_<^{5/3}}{V_>^{2/3}}\,{}_2F_1\left(\frac{2}{3},\frac{7}{3},\frac{8}{3};\frac{V_<}{V_>}\right)+\mathrm{c.c.},\label{Result:S34_LowT}
\end{align}
where we have introduced $V_>\equiv\max(\abs{V_1},\abs{V_2})$ and $V_<\equiv\min(\abs{V_1},\abs{V_2})$ for notational convenience. The low-temperature asymptotic result in Eq.~\eqref{Result:S34_LowT} makes the scaling structure explicit. Within the large-device low-temperature regime, the flux-dependent cross-correlation is linear in the overall bias scale $V_>$, while the bias mismatch enters only through the dimensionless ratio $r\equiv V_</V_>$. In particular, in the strongly asymmetric limit $r\ll1$, one finds $S_{34}^\Phi\propto V_>\,r^{5/3}$, since ${}_2F_1(2/3,7/3,8/3;r)\to1$ as $r\to0$. On the other hand, the symmetric bias limit $r\to1$ is enhanced, with the apparent singular behavior rounded by finite-temperature effects beyond the strict asymptotic form. This behavior is illustrated in Fig.~\ref{fig:S34_ratio}. The derivation of the asymptotic kernel and the hypergeometric reduction is given in Appendix~\ref{Appendix:low-T_limit}.

\subsection{Temperature dependence and thermal crossover}

\begin{figure}
\centering
\includegraphics[width=0.99\linewidth]{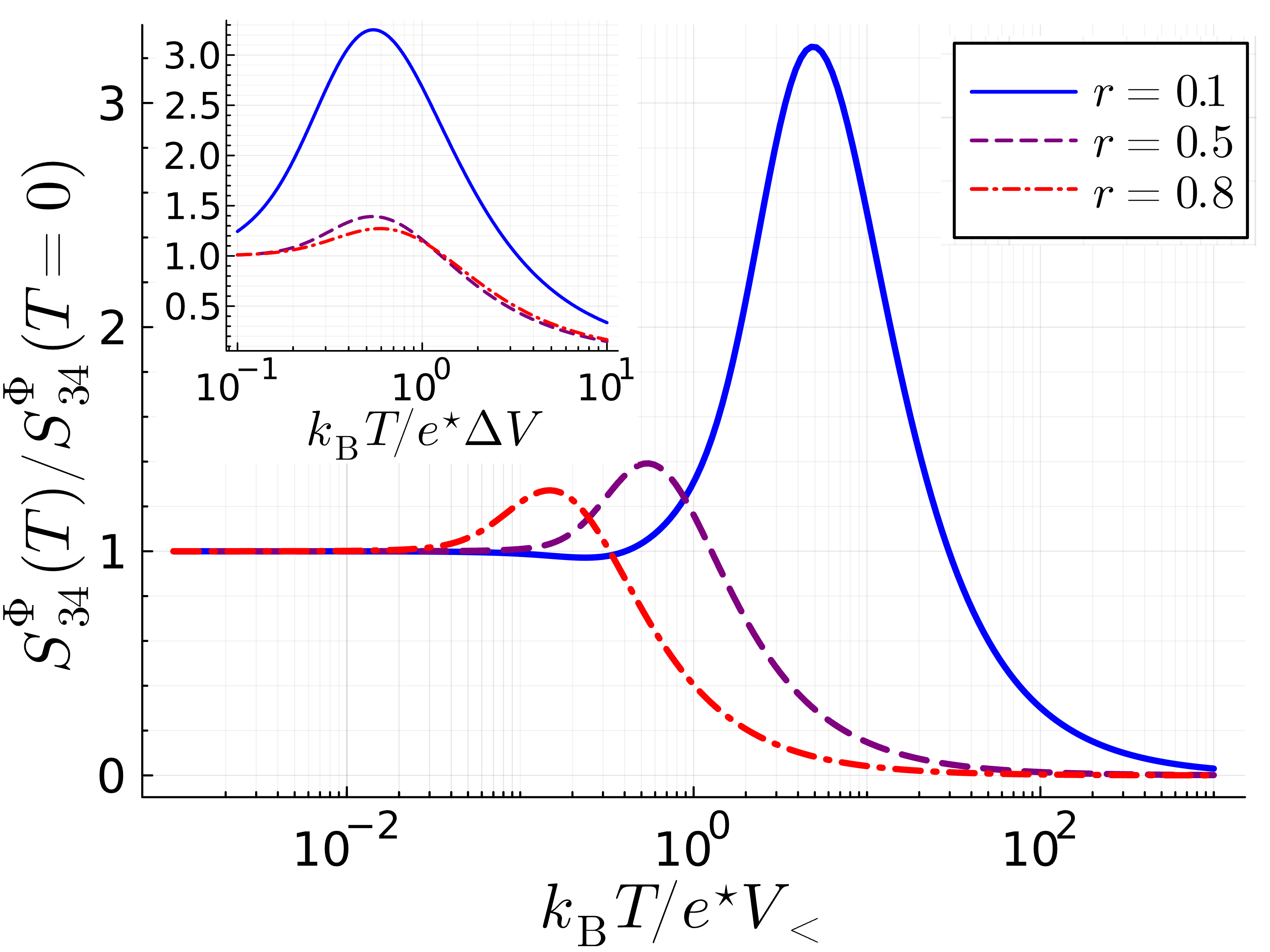}
\caption{Temperature dependence of the flux-dependent cross-correlation for several fixed values of the bias mismatch ratio $r{=V_</V_>}$, shown in normalized form as a function of $k_\mathrm{B}T/e^\star V_<$ at $\Delta\tau=0$. The inset shows the same data replotted against $k_\mathrm{B}T/e^\star\Delta V$ with $\Delta V\equiv V_>-V_<$ to highlight the {effect} of the bias mismatch.}
\label{fig:S34_T}
\end{figure}

To examine the thermal crossover away from the symmetric-bias point, we plot the normalized interference signal for fixed $r<1$ as a function of $k_\mathrm{B}T/e^\star V_<$ in Fig.~\ref{fig:S34_T}. This representation expresses the temperature relative to the smaller of the two bias scales, $V_<$, which sets the boundary of the zero-temperature frequency window. This, however, does not mean that the finite-temperature crossover is governed by $V_<$ alone. At $T=0$, the smaller bias $V_<$ determines the boundary of the allowed frequency range, but the effective spectral overlap entering the interference term is additionally suppressed by the bias mismatch $\Delta V\equiv V_>-V_<$. Finite temperature broadens the sharp spectral edges and partially restores the overlap lost at zero temperature. As a result, the initial enhancement of the interference signal sets in only once thermal broadening becomes comparable to the mismatch-induced phase-space restriction. This explains why the onset is delayed for smaller $r$, for which $\Delta V=V_>(1-r)$ is larger. At the same time, the full crossover profile remains strongly $r$-dependent, since the kernel retains a nontrivial dependence on the bias ratio even within the zero-temperature support window. For sufficiently strong bias mismatch, the temperature dependence can become nonmonotonic. At intermediate temperature{s}, thermal broadening enhances the signal by relaxing the mismatch constraint, whereas at high temperature{s}, the interference is suppressed by thermal decoherence. This interpretation is consistent with the inset of Fig.~\ref{fig:S34_T}, where the same data are replotted against $k_\mathrm{B}T/e^\star\Delta V$, highlighting its role in the delayed thermal onset.

\subsection{Role of the arm mismatch $\Delta\tau$}

\begin{figure}
\centering
\includegraphics[width=0.99\linewidth]{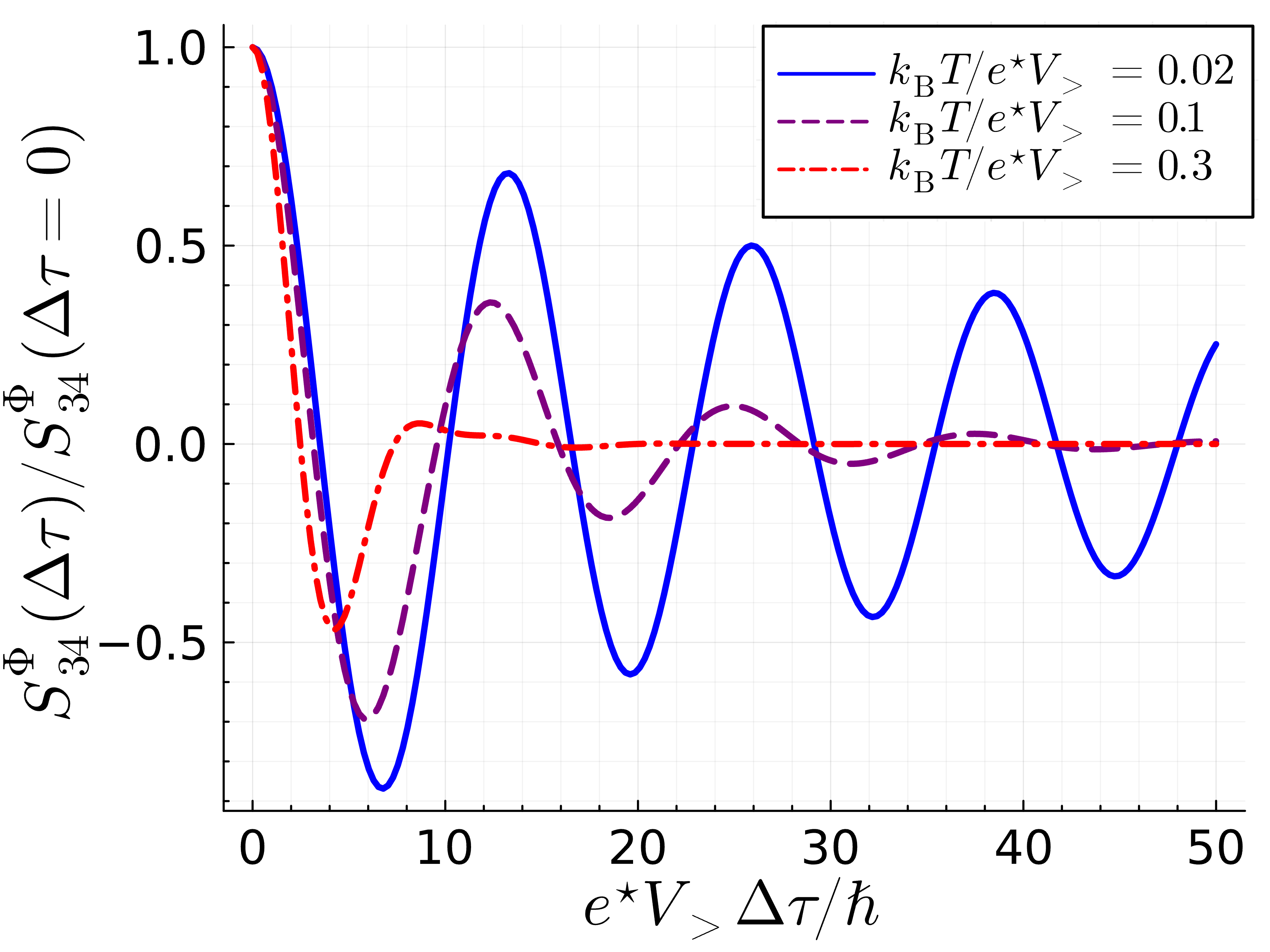}
\caption{The normalized flux-dependent cross-correlation as a function of the dimensionless arm-mismatch parameter $e^\star V_>\Delta\tau/\hbar$, evaluated at $r{=V_</V_>}=0.5$. As the arm mismatch increases, the interference contribution exhibits damped oscillations, while increasing temperature progressively suppresses the oscillation amplitude.}
\label{fig:S34_tau}
\end{figure}

Having established the behavior of $S_{34}^{\Phi}$ in the matched-arm limit $\Delta\tau=0$, we now turn to the effect of a finite propagation-time mismatch between the two interfering paths. While the $\Delta\tau=0$ result is particularly useful for identifying the bias and temperature scaling of the interference contribution, a nonzero $\Delta\tau$ provides an additional control parameter that directly probes its frequency structure.
Beyond the matched-arm limit, the cross-correlation $S_{34}^{\Phi}$ acquires a nontrivial dependence on the arm mismatch through the oscillating factor $e^{i\omega\Delta\tau}$ in the frequency integral. In the perfectly matched case, $\Delta\tau=0$, the integrand is strictly positive and the visibility of the AB oscillation is maximized. For $\Delta\tau \neq 0$, by contrast, the oscillatory phase factor leads to partial cancellations among different frequency components, thereby reducing the visibility and modifying the interference pattern. At sufficiently high temperatures, however, the thermal envelope suppresses the interference before the first zero in $\Delta\tau$ is reached, so that the arm-mismatch dependence becomes effectively monotonic rather than visibly oscillatory. As shown in Fig.~\ref{fig:S34_tau}, the arm-mismatch dependence exhibits damped oscillations as a function of the dimensionless mismatch parameter $e^\star V_{>}\Delta\tau/\hbar$, while thermal decoherence progressively suppresses their amplitude. The arm mismatch thus serves as an experimentally tunable knob: by varying the gate-defined path lengths, one can probe the Fourier structure, whose frequency dependence is governed by the scaling dimensions of the tunneling operators.

\begin{figure*}
\centering
\includegraphics[width=0.99\linewidth]{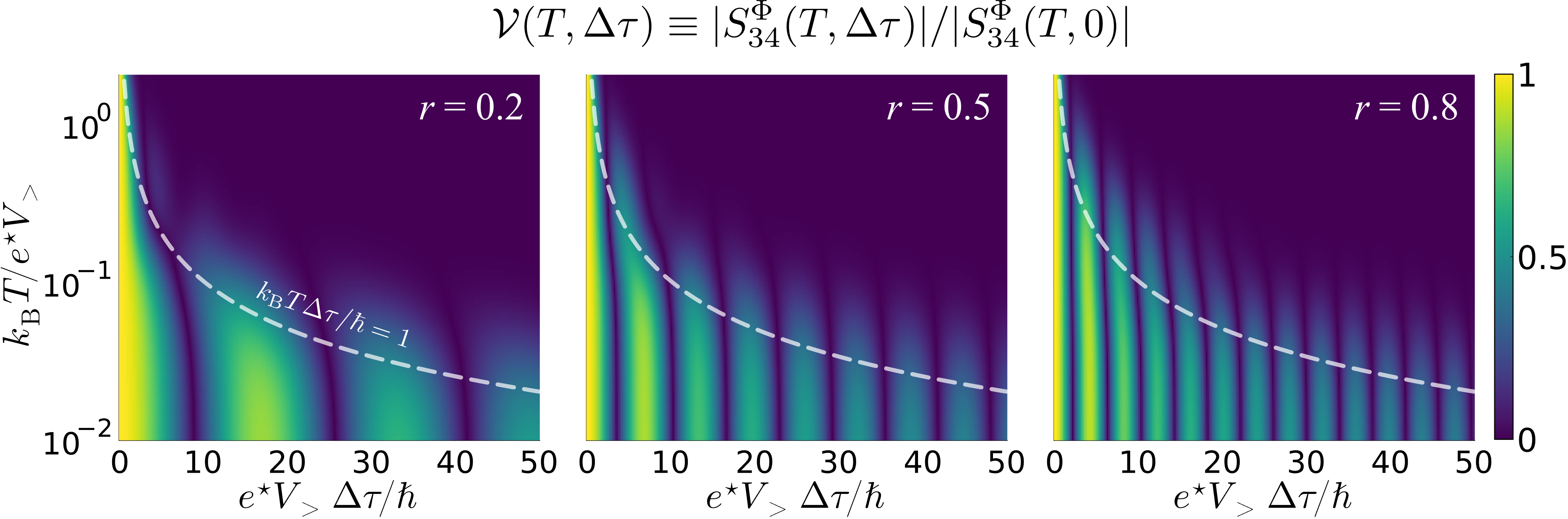}
\caption{Visibility of the flux-dependent cross-correlation, defined as $\mathcal{V}(T,\Delta\tau)\equiv\abs{S_{34}^\Phi(T,\Delta\tau)}/\abs{S_{34}^\Phi(T,0)}$, plotted as a function of the dimensionless arm-mismatch parameter $e^\star V_>\Delta\tau/\hbar$ and the reduced temperature $k_\mathrm{B}T/e^\star V_>$ {for several fixed values of the bias mismatch ratio $r=V_</V_>$}. The dashed line indicates $k_\mathrm{B}T\Delta\tau/\hbar=1$, which marks the crossover beyond which the interference is substantially suppressed.}
\label{fig:visibility}
\end{figure*}

%[\#2 Add comments on noise visibility]
To characterize more systematically the interplay between finite temperature and propagation-time mismatch, we introduce a visibility associated with the flux-dependent part of the cross-correlation,
\begin{equation}
    \mathcal{V}(T,\Delta\tau)\equiv\frac{\abs{S_{34}^\Phi(T,\Delta\tau)}}{\abs{S_{34}^\Phi(T,0)}}.
\end{equation}
Fig.~\ref{fig:visibility} shows this quantity as a function of the dimensionless mismatch $e^\star V_>\Delta\tau/\hbar$ and the reduced temperature $k_\mathrm{B}T/e^\star V_>$. This representation makes {it} clear that the suppression of the interference is controlled not by temperature or arm mismatch separately, but by their combined effect. In particular, the visibility begins to decrease appreciably once the product $k_\mathrm{B}T\Delta\tau/\hbar$ becomes of order unity~\cite{Chung2005PRB}. 
The visibility map thus provides a compact way of visualizing the thermal coherence constraint associated with arm mismatch in the present fractional HBT geometry.

\subsection{Flux-independent background}

In addition to the AB-sensitive contribution $S_{34}^{\Phi}$, Eq.~\eqref{Eq:S34} generates terms involving at most two QPCs that are independent of the enclosed flux. These terms contribute a flux-independent background to $S_{34}$, which is of the same perturbative order in the tunneling amplitudes. In experiment{s}, $S_{34}^{\Phi}$ can be isolated from this background by modulating the AB flux and extracting the oscillating component.
    
\subsection{Experimental considerations}

The proposed device requires a $\nu=2/5$ fractional quantum Hall state with two co-propagating edge modes that can be brought into close proximity at four separate constrictions. Such multi-constriction geometries have been realized in GaAs/AlGaAs heterostructures for integer and fractional quantum Hall edge {states}~\cite{Henny1999Science,Bartolomei2020Science}, and similar fabrication strategies may be applicable to the $\nu=2/5$ regime. The large-device condition $L_{\alpha}/v_{\alpha}\gg\beta\hbar$ requires the propagation time between neighboring QPCs to exceed the thermal coherence time $\hbar/k_{B}T$. For typical edge velocities $v\sim 10^{4}$--$10^{5}\,{\rm m}/{\rm s}$ and temperatures $T\sim 20\,{\rm mK}$, this corresponds to characteristic length scales in the several-$\mu$m to several-tens-of-$\mu$m range, depending on the edge velocity and on how {strictly} the large-device condition is imposed. These scales remain broadly compatible with present lithographic capabilities and are comparable to the size of the previous electron two-particle interferometer~\cite{Neder2007Nature}, although a more detailed device-specific estimate would be needed for a concrete implementation. 

%[\#2]
As summarized by the visibility map in Fig.~\ref{fig:visibility}, the experimental requirements can be expressed in terms of three relevant time scales: the propagation-time mismatch $\Delta\tau$ between the two interfering two-particle {paths}, the thermal coherence time $\hbar/k_\mathrm{B}T$, and the edge-mode times of flight $L_\alpha/v_\alpha$ between neighboring QPCs. {To} retain a substantial flux-dependent signal, the mismatch must remain shorter than the thermal coherence time, $\Delta\tau\lesssim\hbar/k_\mathrm{B}T$. Combined with the large-device condition, this {can} be summarized as
\begin{equation}
\Delta\tau\lesssim\frac{\hbar}{k_\mathrm{B}T}\ll\frac{L_\alpha}{v_\alpha}.
\end{equation}
For a representative operating temperature of $T\sim20\,\mathrm{mK}$, one finds $\Delta\tau\lesssim0.4\,\mathrm{ns}$, up to a factor of order unity.

\subsection{Beyond the large-device limit}

As emphasized above, the cancellation of anyonic exchange phases in the present analysis relies on the temporal separation induced by the large-device limit, which renders the weighted Keldysh sums insensitive to contour orderings (see Appendix~\ref{All-contour}). When the device size becomes comparable to the thermal length, $L_{\alpha}/v_{\alpha}\sim \beta\hbar$, this temporal separation {is reduced} and the cancelation mechanism may no longer persist. In that regime, contributions carrying statistical phases such as $1+e^{\pm i\pi/3}$ may survive in the cross-correlation, so that the interference pattern could acquire a more direct dependence on the anyonic statistical angle, thereby providing a possible route toward a more direct probe of the exchange phase from two-particle interference. Clarifying the quantitative behavior in this crossover regime is an interesting direction for future work.

\section{Summary}
\label{sec:summary} 

In this paper, we have investigated a Hanbury Brown{--}Twiss interferometer for a $\nu=2/5$ fractional quantum Hall edge with four weak tunneling constrictions. Using a bosonized edge theory and a perturbative Keldysh analysis, we derived the flux-dependent part of the zero-frequency current cross-correlation. In the large-device limit, the result reduces to a compact expression with the same overall structure as in the electronic case, but governed by the fractional quasiparticle charge and by the scaling dimensions of the fractional edge modes. Our analysis also shows that the contour orderings carrying explicit anyonic exchange phases cancel in this limit, so that the surviving interference signal reflects the fractional nature of the tunneling quasiparticles primarily through their charge and edge scaling properties.

Our results demonstrate that HBT interferometry can be naturally implemented in hierarchical fractional quantum Hall edges with multiple co-propagating modes. The proposed geometry provides a controlled platform for studying two-particle interference of fractional quasiparticles. While the explicit anyonic exchange phases cancel in the large-device limit analyzed here, our analysis indicates that finite-size devices offer a promising route to directly access the anyonic statistical angle through current-noise measurements.

\textit{Note added.} While preparing this manuscript, we became aware of a related work~\cite{puster2026arXiv}.

\begin{acknowledgments}
This work was supported by {JSPS KAKENHI (Grant No. JP25H00613, JP24H00827, JP24K06951, JP22H00112)} from MEXT, Japan and the Japan Science and Technology Agency (JST) ASPIRE Program No. JPMJAP2410. This work was also supported by Hirose Foundation, {JST SPRING (Grant No.~JPMJSP2108),} and Materials Education program for the Innovation of Technology (MERIT), the University of Tokyo. This work was carried out in the framework of the
project ``ANY-HALL'' (Grant No. ANR-21-CE30-0064-03). We acknowledge
funding from the Agence Nationale de la Recherche under
the France 2030 programme, reference ANR-22-PETQ-0012, and from the ERC advanced grant
``ASTEC'' (Grant No. 101096610).
\end{acknowledgments}

\appendix

\section{Keldysh contour orderings and Klein-factor phases}
\label{All-contour} 

In this Appendix, we enumerate the AB-sensitive second-order terms generated by contour ordering of the four tunneling operators $\{A_{1},A_{3},A_{2}^{\dagger},A_{4}^{\dagger}\}$. There are eight inequivalent contour orderings {that contribute at leading order to} $\Gamma_{1}\Gamma_{2}^{\ast}\Gamma_{3}\Gamma_{4}^{\ast}$. In the expressions below, we suppress overall prefactors common to all AB-sensitive orderings, including $\Gamma_{1}\Gamma_{2}^{\ast}\Gamma_{3}\Gamma_{4}^{\ast}/(2\pi)^{4}(a_{1/3}a_{1/15})^{2}$ and the AB phase $e^{2\pi i\Phi/\Phi^\star_{{0}}}$. For each ordering, we list (i) the Klein-factor product in the corresponding operator order and (ii) the resulting product of vertex propagators $\mathcal{G}^{\eta\eta'}_{\alpha}$ with the appropriate time arguments and coordinate signs. The complex-conjugate contribution follows from $A_{i}\leftrightarrow A_{i}^{\dagger}$, $\Gamma_{i}\leftrightarrow \Gamma_{i}^{\ast}$, and $\Phi\leftrightarrow -\Phi$. At order $\Gamma_{1}\Gamma_{2}^{\ast} \Gamma_{3}\Gamma_{4}^{\ast}$, the AB-sensitive part of the contour-ordered average can be organized into the following eight operator orderings:
\begin{widetext}
\begin{align}
 & \expval{A_1(t_\eta)A_3(t'_{-\eta})A_2^\dagger(t_{1,\eta_1}) A_4^\dagger(t_{2,\eta_2})}_{0}+\expval{A_1(t_\eta)A_3(t'_{-\eta})A_4^\dagger(t_{1,\eta_1}) A_2^\dagger(t_{2,\eta_2})}_{0} \\
 & +\expval{A_2^\dagger(t_\eta)A_4^\dagger(t'_{-\eta})A_1(t_{1,\eta_1})A_3(t_{2,\eta_2})}_{0}+\expval{A_2^\dagger(t_\eta)A_4^\dagger(t'_{-\eta})A_3(t_{1,\eta_1})A_1(t_{2,\eta_2})}_{0}  \\
 & -\expval{A_1(t_\eta)A_4^\dagger(t'_{-\eta})A_2^\dagger(t_{1,\eta_1})A_3(t_{2,\eta_2})}_{0}-\expval{A_1(t_\eta)A_4^\dagger(t'_{-\eta})A_3(t_{1,\eta_1})A_2^\dagger(t_{2,\eta_2})}_{0}  \\
 & -\expval{A_2^\dagger(t_\eta)A_3(t'_{-\eta})A_1(t_{1,\eta_1})A_4^\dagger(t_{2,\eta_2})}_{0}-\expval{A_2^\dagger(t_\eta)A_3(t'_{-\eta})A_4^\dagger(t_{1,\eta_1})A_1(t_{2,\eta_2})}_{0}.
\end{align}
Evaluating each term gives the following Klein-factor products and vertex-propagator structures:
\begin{align}
\mathcal{I} & \equiv\mathcal{F}_{1}\mathcal{F}_{3}\mathcal{F}_{2}^{\dagger}\mathcal{F}_{4}^{\dagger}\mathcal{G}_{\mathrm{L}}^{\eta\eta_1}(t-t_{1}+L_{\mathrm{L}}/v_{1/15})\mathcal{G}_{\mathrm{U}}^{\eta\eta_2}(t-t_{2}+L_{\mathrm{U}}/v_{1/3})\mathcal{G}_{\mathrm{R}}^{-\eta\eta_2}(t'-t_{2}+L_{\mathrm{R}}/v_{1/15})\mathcal{G}_{\mathrm{D}}^{-\eta\eta_1}(t'-t_{1}+L_{\mathrm{D}}/v_{1/3})\nonumber \\
& \quad+\mathcal{F}_{1}\mathcal{F}_{3}\mathcal{F}_{4}^{\dagger}\mathcal{F}_{2}^{\dagger}\mathcal{G}_{\mathrm{L}}^{\eta\eta_2}(t-t_{2}+L_{\mathrm{L}}/v_{1/15})\mathcal{G}_{\mathrm{U}}^{\eta\eta_1}(t-t_{1}+L_{\mathrm{U}}/v_{1/3})\mathcal{G}_{\mathrm{R}}^{-\eta\eta_1}(t'-t_{1}+L_{\mathrm{R}}/v_{1/15})\mathcal{G}_{\mathrm{D}}^{-\eta\eta_2}(t'-t_{2}+L_{\mathrm{D}}/v_{1/3})          \\
& +\mathcal{F}_{2}^{\dagger}\mathcal{F}_{4}^{\dagger}\mathcal{F}_{1}\mathcal{F}_{3}\mathcal{G}_{\mathrm{D}}^{\eta\eta_2}(t-t_{2}-L_{\mathrm{D}}/v_{1/3})\mathcal{G}_{\mathrm{L}}^{\eta\eta_1}(t-t_{1}-L_{\mathrm{L}}/v_{1/15})\mathcal{G}_{\mathrm{U}}^{-\eta\eta_1}(t'-t_{1}-L_{\mathrm{U}}/v_{1/3})\mathcal{G}_{\mathrm{R}}^{-\eta\eta_2}(t'-t_{2}-L_{\mathrm{R}}/v_{1/15})\nonumber      \\
& \quad+\mathcal{F}_{2}^{\dagger}\mathcal{F}_{4}^{\dagger}\mathcal{F}_{3}\mathcal{F}_{1}\mathcal{G}_{\mathrm{D}}^{\eta\eta_1}(t-t_{1}-L_{\mathrm{D}}/v_{1/3})\mathcal{G}_{\mathrm{L}}^{\eta\eta_2}(t-t_{2}-L_{\mathrm{L}}/v_{1/15})\mathcal{G}_{\mathrm{U}}^{-\eta\eta_2}(t'-t_{2}-L_{\mathrm{U}}/v_{1/3})\mathcal{G}_{\mathrm{R}}^{-\eta\eta_1}(t'-t_{1}-L_{\mathrm{R}}/v_{1/15})          \\
& -\mathcal{F}_{1}\mathcal{F}_{4}^{\dagger}\mathcal{F}_{2}^{\dagger}\mathcal{F}_{3}\mathcal{G}_{\mathrm{L}}^{\eta\eta_1}(t-t_{1}+L_{\mathrm{L}}/v_{1/15})\mathcal{G}_{\mathrm{U}}^{\eta-\eta}(t-t'+L_{\mathrm{U}}/v_{1/3})\mathcal{G}_{\mathrm{R}}^{-\eta\eta_2}(t'-t_{2}-L_{\mathrm{R}}/v_{1/15})\mathcal{G}_{\mathrm{D}}^{\eta_1\eta_2}(t_{1}-t_{2}-L_{\mathrm{D}}/v_{1/3})\nonumber      \\
& \quad-\mathcal{F}_{1}\mathcal{F}_{4}^{\dagger}\mathcal{F}_{3}\mathcal{F}_{2}^{\dagger}\mathcal{G}_{\mathrm{L}}^{\eta\eta_2}(t-t_{2}+L_{\mathrm{L}}/v_{1/15})\mathcal{G}_{\mathrm{U}}^{\eta-\eta}(t-t'+L_{\mathrm{U}}/v_{1/3})\mathcal{G}_{\mathrm{R}}^{-\eta\eta_1}(t'-t_{1}-L_{\mathrm{R}}/v_{1/15})\mathcal{G}_{\mathrm{D}}^{\eta_1\eta_2}(t_{1}-t_{2}+L_{\mathrm{D}}/v_{1/3})          \\
& -\mathcal{F}_{2}^{\dagger}\mathcal{F}_{3}\mathcal{F}_{1}\mathcal{F}_{4}^{\dagger}\mathcal{G}_{\mathrm{D}}^{\eta-\eta}(t-t'-L_{\mathrm{D}}/v_{1/3})\mathcal{G}_{\mathrm{L}}^{\eta\eta_2}(t-t_{2}-L_{\mathrm{L}}/v_{1/15})\mathcal{G}_{\mathrm{R}}^{-\eta\eta_1}(t'-t_{1}+L_{\mathrm{R}}/v_{1/15})\mathcal{G}_{\mathrm{U}}^{\eta_1\eta_2}(t_{1}-t_{2}-L_{\mathrm{U}}/v_{1/3})\nonumber      \\
& \quad-\mathcal{F}_{2}^{\dagger}\mathcal{F}_{3}\mathcal{F}_{4}^{\dagger}\mathcal{F}_{1}\mathcal{G}_{\mathrm{D}}^{\eta-\eta}(t-t'-L_{\mathrm{D}}/v_{1/3})\mathcal{G}_{\mathrm{L}}^{\eta\eta_1}(t-t_{1}-L_{\mathrm{L}}/v_{1/15})\mathcal{G}_{\mathrm{R}}^{-\eta\eta_2}(t'-t_{2}+L_{\mathrm{R}}/v_{1/15})\mathcal{G}_{\mathrm{U}}^{\eta_1\eta_2}(t_{1}-t_{2}+L_{\mathrm{U}}/v_{1/3}).
\end{align}
For brevity, we have also omitted the bias-dependent dynamical phase factors absorbed into the shifted time variables defined in Eq.~\eqref{Eq:bias-shift}. Since $(t_{1},\eta_{1})$ and $(t_{2},\eta_{2})$ are dummy integration variables, relabeling $(t_{1},\eta_{1})\leftrightarrow(t_{2},\eta_{2})$ maps each paired ordering into its partner and leaves the bosonic vertex part invariant. Including the Klein factors, we can therefore group the eight AB-sensitive terms into four paired contributions, since they may differ in the ordering of the Klein factors, which can generate anyonic statistical phases when neighboring Klein factors are exchanged under the cyclic algebra. {U}sing the identity $\mathcal{G}^{\eta\eta'}_{\alpha}(\tau)=\mathcal{G}^{\eta'\eta}_{\alpha}(-\tau)$, we arrive at $\mathcal{I}=\mathcal{I}_{1}+\mathcal{I}_{2}+\mathcal{I}_{3}+\mathcal{I}_{4}$ with
\begin{align}
\mathcal{I}_{1} & =2\mathcal{G}_{\mathrm{L}}^{\eta_1\eta}(t_{1}-t-L_{\mathrm{L}}/v_{1/15})\mathcal{G}_{\mathrm{U}}^{\eta_2\eta}(t_{2}-t-L_{\mathrm{U}}/v_{1/3})\mathcal{G}_{\mathrm{R}}^{\eta_2-\eta}(t_{2}-t'-L_{\mathrm{R}}/v_{1/15})\mathcal{G}_{\mathrm{D}}^{\eta_1-\eta}(t_{1}-t'-L_{\mathrm{D}}/v_{1/3}),\label{A1}  \\
\mathcal{I}_{2} & =2\mathcal{G}_{\mathrm{D}}^{\eta\eta_2}(t-t_{2}-L_{\mathrm{D}}/v_{1/3})\mathcal{G}_{\mathrm{L}}^{\eta\eta_1}(t-t_{1}-L_{\mathrm{L}}/v_{1/15})\mathcal{G}_{\mathrm{U}}^{-\eta\eta_1}(t'-t_{1}-L_{\mathrm{U}}/v_{1/3})\mathcal{G}_{\mathrm{R}}^{-\eta\eta_2}(t'-t_{2}-L_{\mathrm{R}}/v_{1/15}),\label{A2} \\
\mathcal{I}_{3} & =-\bigl(1+e^{i\pi/3}\bigr)\mathcal{G}_{\mathrm{L}}^{\eta_1\eta}(t_{1}-t-L_{\mathrm{L}}/v_{1/15})\mathcal{G}_{\mathrm{U}}^{-\eta\eta}(t'-t-L_{\mathrm{U}}/v_{1/3})\mathcal{G}_{\mathrm{R}}^{-\eta\eta_2}(t'-t_{2}-L_{\mathrm{R}}/v_{1/15})\mathcal{G}_{\mathrm{D}}^{\eta_1\eta_2}(t_{1}-t_{2}-L_{\mathrm{D}}/v_{1/3}),\label{A3}  \\
\mathcal{I}_{4} & =-\bigl(1+e^{-i\pi/3}\bigr)\mathcal{G}_{\mathrm{D}}^{\eta-\eta}(t-t'-L_{\mathrm{D}}/v_{1/3})\mathcal{G}_{\mathrm{L}}^{\eta\eta_2}(t-t_{2}-L_{\mathrm{L}}/v_{1/15})\mathcal{G}_{\mathrm{R}}^{\eta_1-\eta}(t_{1}-t'-L_{\mathrm{R}}/v_{1/15})\mathcal{G}_{\mathrm{U}}^{\eta_1\eta_2}(t_{1}-t_{2}-L_{\mathrm{U}}/v_{1/3})\label{A4}.
\end{align}
The factors 2 in Eqs.~\eqref{A1} and \eqref{A2} follow because the paired orderings are related by $(t_{1},\eta_{1})\leftrightarrow(t_{2},\eta_{2})$ and the exchanged Klein factors are non-neighboring. The factors $1+e^{\pm i\pi/3}$ in Eqs.~\eqref{A3} and \eqref{A4} arise from the cyclic exchange algebra Eq.~\eqref{Eq:Klein-algebra} when neighboring Klein factors are permuted. In the large{-}device limit, the vertex propagator becomes independent of the first Keldysh index. Consequently, the contributions $\mathcal{I}_{1}$, $\mathcal{I}_{3}$, and $\mathcal{I}_{4}$ vanish upon the weighted Keldysh sums over $\eta_{1}$ and $\eta_{2}$, whereas $\mathcal{I}_{2}$ survives and yields Eq.~\eqref{Eq:gggg}.

It is worth emphasizing that the explicit statistical phase factors $1+e^{\pm i\pi/3}$ appear only in $\mathcal{I}_{3}$ and $\mathcal{I}_{4}$. In the large-device limit, precisely these contributions vanish after the weighted Keldysh sums are carried out. Thus, {the flux-dependent contribution} in the main text no longer {depends on} the anyonic statistical angle.

\section{Derivation of $g_{\alpha}^{\eta\eta'}(\omega)$ in the large-device limit}
\label{g_omega} 
        
In the large{-}device limit, the vertex propagator $g_{\alpha}^{\eta\eta'}(\omega)$ depends only on the second Keldysh index $\eta'$, and Eq.~\eqref{def:Fourier} reduces to
\begin{align}
g^{\eta\eta'}_{\alpha}(\omega)\xrightarrow{L_\alpha/v_\alpha\gg\beta\hbar}\int_{-\infty}^{+\infty}\dd{\tau}e^{i\omega\tau}\left[\frac{\sinh\frac{i\pi\tau_{0}}{\beta\hbar}}{\sinh\frac{\pi}{\beta\hbar}\left(i\tau_{0}-\eta'\tau\right)}\right]^{\lambda_\alpha^2\nu_\alpha}.
\end{align}
Splitting the integration range into $\tau>0$ and $\tau<0$, we write
\begin{equation}
\int_{-\infty}^{+\infty} \dd{\tau} e^{i\omega\tau} \left[\frac{\sinh\frac{i\pi\tau_{0}}{\beta\hbar}}{\sinh\frac{\pi}{\beta\hbar}\left(i\tau_{0}-\eta'\tau\right)}
\right]^{\lambda_\alpha^2\nu_\alpha}=\sum_{\epsilon=\pm}\int_{0}^{\infty} \dd{\tau} e^{i\epsilon\omega\tau} \left[\frac{\sinh\frac{i\pi\tau_{0}}{\beta\hbar}}{\sinh\frac{\pi}{\beta\hbar}\left(i\tau_{0}-\epsilon\eta'\tau\right)}\right]^{\lambda_\alpha^2\nu_\alpha}
\end{equation}
and we obtain the following form
\begin{align}
& \int_{0}^{\infty} \dd{\tau} e^{i\epsilon\omega\tau}\left[\frac{\sinh\frac{i\pi\tau_{0}}{\beta\hbar}}{\sinh\frac{\pi}{\beta\hbar}\left(i\tau_{0}-\epsilon\eta'\tau\right)}\right]^{\lambda_\alpha^2\nu_\alpha}\nonumber \\
& =\frac{\beta\hbar}{2\pi}\left(\frac{2\pi\tau_{0}}{\beta\hbar}\right)^{\lambda_\alpha^2\nu_\alpha}e^{-i\epsilon\eta'\pi\lambda_\alpha^2\nu_\alpha/2}B\left(\frac{\lambda_{\alpha}^{2}\nu_{\alpha}}{2}-i\epsilon\frac{\beta\hbar\omega}{2\pi},1\right){}_{2}F_{1}\left(\lambda_{\alpha}^{2}\nu_{\alpha},\frac{\lambda_{\alpha}^{2}\nu_{\alpha}}{2}-i\epsilon\frac{\beta\hbar\omega}{2\pi},1+\frac{\lambda_{\alpha}^{2}\nu_{\alpha}}{2}- i\epsilon\frac{\beta\hbar\omega}{2\pi};e^{2\pi i\epsilon\eta'\tau_0/\beta\hbar}\right).\label{B3}
\end{align}
Taking the limit $\tau_{0}\to0^{+}$ {requires some care when $\lambda_\alpha^2\nu_\alpha>1$, since} the standard identity {for} the hypergeometric function
\begin{equation}
{}_{2}F_{1}(a,b,c;1)=\frac{\Gamma(c)\Gamma(c-a-b)}{\Gamma(c-a)\Gamma(c-b)}
,\qquad{{\Re(c-a-b)>0},}\label{B4}
\end{equation}
{is then not directly applicable. We therefore analyze Eq.~\eqref{B3} using the connection formula around $z=1$:
\begin{align}
{}_2F_1(a,b,c;z)&=\frac{\Gamma(c)\Gamma(c-a-b)}{\Gamma(c-a)\Gamma(c-b)}{}_2F_1(a,b,a+b-c+1;1-z)\nonumber\\
&\qquad+\frac{\Gamma(c)\Gamma(a+b-c)}{\Gamma(a)\Gamma(b)}(1-z)^{c-a-b}{}_2F_1(c-a,c-b,c-a-b+1;1-z).\label{B5}
\end{align}
Applying Eq.~\eqref{B5} to Eq.~\eqref{B3}, we decompose the result into a Gauss term and a branch term. The Gauss term gives the beta-function contribution obtained below. For the branch term, we use
\begin{equation}
1-e^{2\pi i\epsilon\eta'\tau_0/\beta\hbar}=-i\epsilon\eta'\frac{2\pi\tau_0}{\beta\hbar}+O(\tau_0^2),
\end{equation}
so that the branch factor $(1-z)^{c-a-b}$ behaves on the principal branch as
\begin{equation}
(1-e^{2\pi i\epsilon\eta'\tau_0/\beta\hbar})^{1-\lambda_\alpha^2\nu_\alpha}=\left(\frac{2\pi\tau_0}{\beta\hbar}\right)^{1-\lambda_\alpha^2\nu_\alpha}e^{-i\pi\epsilon\eta'(1-\lambda_\alpha^2\nu_\alpha)/2}[1+O(\tau_0)].
\end{equation}
After restoring the prefactor in Eq.~\eqref{B3}, the branch contribution is, to leading order in $\tau_0$, given by
\begin{equation}
\frac{\tau_0}{\lambda_\alpha^2\nu_\alpha-1}e^{-i\epsilon\eta'\pi/2}+O(\tau_0^2),
\end{equation}
where we have used $B(1,\lambda_\alpha^2\nu_\alpha-1) = (\lambda_\alpha^2\nu_\alpha-1)^{-1}$.
%up to the order of $\tau_0$. 
Summing over $\epsilon=\pm$, we find
\begin{equation}
\sum_{\epsilon=\pm}e^{-i\epsilon\eta'\pi/2}=e^{-i\eta'\pi/2}+e^{i\eta'\pi/2}=0.
\end{equation}
For the L and R channels, the Gauss part in Eq.~\eqref{B3} contributes at order $O(\tau_0^{5/3})$. By contrast, the remaining branch contribution starts at $O(\tau_0^2)$ and is therefore subleading. Thus, after summing over $\epsilon=\pm$, the $\tau_0\to0^+$ limit is determined solely by the Gauss part, and} we obtain
\begin{equation} 
\sum_{\epsilon=\pm}\int_{0}^{\infty} \dd{\tau} e^{i\epsilon\omega\tau} \left[\frac{\sinh\frac{i\pi\tau_{0}}{\beta\hbar}}{\sinh\frac{\pi}{\beta\hbar}\left(i\tau_{0}-\epsilon\eta'\tau\right)}\right]^{\lambda_\alpha^2\nu_\alpha}
\xrightarrow{\tau_0\to0^+}
\frac{\beta\hbar}{2\pi}
\left(\frac{2\pi\tau_{0}}{\beta\hbar}\right)^{\lambda_\alpha^2\nu_\alpha}\sum_{\epsilon=\pm}
e^{-i \epsilon \eta' \pi \lambda_\alpha^2 \nu_\alpha/2} B\left(\frac{\lambda_{\alpha}^{2}\nu_{\alpha}}{2}
-i\epsilon\frac{\beta\hbar\omega}{2\pi},1-\lambda_{\alpha}^{2}\nu_{\alpha} \right).
\end{equation}
Using the Euler reflection formula
\begin{equation}
\Gamma(z)\Gamma(1-z)=\frac{\pi}{\sin\pi z},
\end{equation}
the beta function can then be rewritten as
\begin{align}
& B\left(\frac{\lambda_{\alpha}^{2}\nu_{\alpha}}{2}-i\epsilon\frac{\beta\hbar\omega}{2\pi},1-\lambda_{\alpha}^{2}\nu_{\alpha}\right)=B\left(\frac{\lambda_{\alpha}^{2}\nu_{\alpha}}{2}-i\epsilon\frac{\beta\hbar\omega}{2\pi},\frac{\lambda_{\alpha}^{2}\nu_{\alpha}}{2}+ i\epsilon\frac{\beta\hbar\omega}{2\pi}\right)\frac{\sin\pi\left(\frac{\lambda_{\alpha}^{2}\nu_{\alpha}}{2}+ i\epsilon\frac{\beta\hbar\omega}{2\pi}\right)}{\sin\pi\lambda_{\alpha}^{2}\nu_{\alpha}}.
\end{align}
Substituting this back and summing over $\epsilon=\pm$, we obtain
\begin{equation}
\sum_{\epsilon=\pm}e^{-i\epsilon\eta'\pi\lambda_\alpha^2\nu_\alpha/2} \sin\pi\left(\frac{\lambda_{\alpha}^{2}\nu_{\alpha}}{2}+i\epsilon\frac{\beta\hbar\omega}{2\pi} \right)
=e^{\eta'\beta\hbar\omega/2}\sin\pi\lambda^{2}_{\alpha}\nu_{\alpha} ,
\end{equation}
\end{widetext}
where the last equality follows from $\cosh x+\eta'\sinh x=e^{\eta'x}$. We thus arrive at Eq.~\eqref{Eq:g}.

\section{Low-temperature reduction of the frequency kernel}\label{Appendix:low-T_limit}

In this Appendix, we derive the low-temperature asymptotic form of the frequency kernel entering Eq.~\eqref{Result:S_34}. We define $\Omega_j\equiv e^\star V_j/\hbar$ $(j=1,2)$ and focus on the regime
\begin{equation}
k_\mathrm{B}T\ll\abs{e^\star V_1},\,\abs{e^\star V_2},\,\abs{e^\star V_1-e^\star V_2}.
\end{equation}
The relevant finite-temperature kernel contains the factor
\begin{widetext}
\begin{align}
&\sinh(\frac{\beta e^{\star}V_{1}}{2})\sinh(\frac{\beta e^{\star}V_{2}}{2})\abs{\Gamma\left(\frac{5}{6}-i\frac{\beta\hbar\omega}{2\pi}\right)}^{4}\abs{\Gamma\left(\frac{1}{6}-i\frac{\beta(e^{\star}V_{1}-\hbar\omega)}{2\pi}\right)}^{2}\abs{\Gamma\left(\frac{1}{6}-i\frac{\beta(e^{\star}V_{2}-\hbar\omega)}{2\pi}\right)}^{2}.
\end{align}
\end{widetext}
Using Stirling's asymptotic formula for fixed $x$ and $\abs{y}\gg1$, 
\begin{equation}
\abs{\Gamma(x+iy)}^2\approx2\pi\abs{y}^{2x-1}e^{-\pi\abs{y}},
\end{equation}
together with
\begin{equation}
\sinh(\frac{\beta e^\star V_j}{2})\approx\frac{\mathrm{sgn}(V_j)}{2}e^{\beta e^\star\abs{V_j}/2},
\end{equation}
the leading low-temperature form of this factor becomes
\begin{align}
&\frac{(2\pi)^4}{4}\frac{\mathrm{sgn}(V_1)\,\mathrm{sgn}(V_2)\,\abs{\omega}^{4/3}}{\abs{\Omega_1-\omega}^{2/3}\abs{\Omega_2-\omega}^{2/3}}e^{-\beta\hbar\mathcal{D}(\omega,\Omega_1,\Omega_2)},
\end{align}
with
\begin{align}
\mathcal{D}(\omega,\Omega_1,\Omega_2)&\equiv\abs{\omega}+\frac{\abs{\Omega_1-\omega}}{2}+\frac{\abs{\Omega_2-\omega}}{2}-\frac{\abs{\Omega_1}}{2}-\frac{\abs{\Omega_2}}{2}\nonumber\\
&=\sum_{j=1}^2\frac{\abs{\omega}+\abs{\Omega_j-\omega}-\abs{\Omega_j}}{2}.
\end{align}
The nonnegativity $\mathcal{D}(\omega,\Omega_1,\Omega_2)\geq0$ follows immediately from the triangle inequality. Moreover, $\mathcal{D}(\omega,\Omega_1,\Omega_2)=0$ holds if and only if both triangle inequalities are saturated, namely $\abs{\omega}+\abs{\Omega_j-\omega}=\abs{\Omega_j}$ for $j=1,2$.
Equivalently, $\omega$ must belong simultaneously to the two intervals
\begin{equation}
[\min(0,\Omega_1),\max(0,\Omega_1)]\cap[\min(0,\Omega_2),\max(0,\Omega_2)].
\end{equation}
In particular, this overlap is nonempty only when $V_1$ and $V_2$ have the same sign, $V_1V_2>0$. Therefore, in the only case where a leading low-temperature contribution survives, the prefactor satisfies $\mathrm{sgn}(V_1)\,\mathrm{sgn}(V_2)=1$. By contrast, when $V_1V_2<0$, the overlap is empty and the kernel is exponentially suppressed at low temperatures.
    
A technical point should be emphasized here. Although the exponential factor correctly identifies the support of the leading low-temperature contribution, Stirling’s approximation is not uniform throughout the entire interval. In particular, it breaks down in the thermal boundary layers near $\omega=0$ or $\Omega_j-\omega=0$, where $\beta\hbar\abs{\omega}$ or $\beta\hbar\abs{\Omega_j-\omega}$ is no longer large. Therefore, the power-law form obtained above is valid only in the interior of the interval, away from the endpoints by an energy scale of order $k_\mathrm{B}T$. The endpoint regions are smoothed by finite-temperature effects and require the original finite-temperature expression, Eq.~\eqref{Result:S_34} for a uniform description. With this caveat in mind, we now evaluate the resulting zero-temperature power-law integral, which controls the bulk low-temperature asymptotics.

For $V_1,V_2>0$, the surviving frequency window is
\begin{equation}
    0<\omega<\min(\Omega_1,\Omega_2),
\end{equation}
whereas for $V_1,V_2<0$, it is
\begin{equation}
    \max(\Omega_1,\Omega_2)=-\min(\abs{\Omega_1},\abs{\Omega_2})<\omega<0.
\end{equation}
In both sign choices, the final result for $\Delta\tau=0$ depends only on $\abs{V_1}$ and $\abs{V_2}$, and therefore reduces to the same Gauss hypergeometric expression given in Eq.~\eqref{Result:S34_LowT}.

For definiteness, let us first consider $V_1,V_2>0$. For notational convenience, we define $\Omega_>\equiv\max(\Omega_1,\Omega_2)$ and $\Omega_<\equiv\min(\Omega_1,\Omega_2)$, so that the relevant zero-temperature integral then takes the form
\begin{equation}
\mathcal{I}_0=\int_0^{\Omega_<}\dd{\omega}\omega^{4/3}(\Omega_1-\omega)^{-2/3}(\Omega_2-\omega)^{-2/3},
\end{equation}
which, after the change of variable $\omega=\Omega_<x$, becomes
\begin{align}
\mathcal{I}_0&=\frac{\Omega_<^{5/3}}{\Omega_>^{2/3}}\int_0^1\dd{x}x^{4/3}(1-x)^{-2/3}(1-rx)^{-2/3}\nonumber\\
&=\frac{\Omega_<^{5/3}}{\Omega_>^{2/3}}\frac{\Gamma(7/3)\Gamma(1/3)}{\Gamma(8/3)}{}_2F_1\left(\frac{2}{3},\frac{7}{3},\frac{8}{3};\frac{\Omega_<}{\Omega_>}\right).
\end{align}
The case $V_1,V_2<0$ is treated analogously and yields the same final expression upon replacing $\Omega_j$ by $\abs{\Omega_j}$. Restoring the $\omega$-independent prefactor omitted above and using the identity $\Gamma(z+1)=z\Gamma(z)$, we recover Eq.~\eqref{Result:S34_LowT}.

\bibliographystyle{apsrev4-2}
\bibliography{ref}

\end{document}